%% file: draft.tex
\documentclass[opre]{informs3aa}

\OneAndAHalfSpacedXI 

\usepackage{endnotes}
\usepackage{mathrsfs}
\let\footnote=\endnote

\usepackage{enumitem}
\usepackage{amsmath}
\usepackage{bm}
\usepackage{framed}
\usepackage[outdir=./]{epstopdf}

\usepackage{graphicx}
\usepackage{xcolor,cancel}
\usepackage{tabularx}
\usepackage{multirow}
\usepackage{subfig}

\usepackage{algorithm}
\usepackage{algorithmic}

\usepackage{longtable}
\usepackage{tabularx}
\usepackage{booktabs}
\usepackage{etex}
\usepackage{multirow}
\usepackage{array}

\usepackage{bigstrut}
\usepackage{soul}

\newcolumntype{C}[1]{>{\centering\arraybackslash}p{#1}}

\newcommand{\bE}{\mathbf{E}}

\newcommand{\cE}{\mathcal{E}}

\newcommand{\ucb}{\textsf{UCB-Add-On}}

\usepackage{natbib}
 \bibpunct[, ]{(}{)}{,}{a}{}{,}%
 \def\bibfont{\smalıl}%
\TheoremsNumberedThrough     
\ECRepeatTheorems

\EquationsNumberedThrough    

\MANUSCRIPTNO{} 

\begin{document}




\TITLE{Online Learning and Optimization for
Revenue Management Problems with Add-on Discounts}


\ARTICLEAUTHORS{

\AUTHOR{David Simchi-Levi}
\AFF{Institute for Data, Systems, and Society, \\
Department of Civil \& Environmental Engineering,
and Operations Research Center,\\
Massachusetts Institute of Technology,
Cambridge, MA, 02139,
\EMAIL{dslevi@mit.edu}}

\AUTHOR{Rui Sun}
\AFF{Institute for Data, Systems, and Society,\\
Massachusetts Institute of Technology,
Cambridge, MA, 02139,
\EMAIL{ruisun@mit.edu}}

\AUTHOR{Huanan Zhang}
\AFF{Harold and Inge Marcus Department of Industrial and Manufacturing Engineering, \\
Pennsylvania State University, PA, 16801, USA 
\EMAIL{huz157@psu.edu}}

}

\ABSTRACT{
We study in this paper a revenue management problem with add-on discounts.
The problem is motivated by the practice in the video game industry, where
a retailer offers discounts on selected \emph{supportive} products (e.g. video games) to customers who have also purchased the  \emph{core} products (e.g. video game consoles).
We formulate this problem as an optimization problem to determine the prices of different products and the selection of products with add-on discounts.
To overcome the computational challenge of this optimization problem, we propose an efficient FPTAS algorithm that can solve the problem approximately to any desired accuracy.
Moreover, we consider the revenue management problem in the setting where
the retailer has no prior knowledge of the demand functions of different products.
To resolve this problem, we propose a UCB-based learning algorithm that uses the FPTAS optimization algorithm as a subroutine.
We show that our learning algorithm can converge to the optimal algorithm that has access to the true demand functions,
and we prove that the convergence rate is tight up to a certain logarithmic term.
In addition, we conduct numerical experiments with the real-world transaction data we collect from a popular video gaming brand's online store on Tmall.com.
The experiment results illustrate our learning algorithm's robust performance and fast convergence in various scenarios. We also compare our algorithm with the optimal policy that does not use any add-on discount, and the results show the advantages of using the add-on discount strategy in practice.
}

\KEYWORDS{revenue management, add-on discount, online learning, approximation algorithm}


\maketitle


\input{Intro}
\input{Model}
\input{Optimization}

\input{Learning}

\input{Discussion}
\input{Numerical}
\input{Conclusion}

\section*{Acknowledgment}
The numerical experiments were done when the second author interned at the Alibaba DAMO Academy of Alibaba Group (US) Inc. under the supervision of Dr. Xinshang Wang and Prof. Wotao Yin.
The authors gratefully acknowledge the support of Dr. Wang and Prof. Yin during the design of the experiments and the revision of the paper.

\def\bibfont{\small}
\bibliographystyle{ormsv080} 
\bibliography{references}

\input{Appendix}

\end{document}

%% file: Intro.tex
\section{Introduction} \label{sec_intro}

The video game industry has been growing fast and steadily in the past two decades.
According to \cite{VBnews2018}, in 2018, the U.S. video game industry matches that of the U.S. film industry on basis of revenue, making around 43 billion USD,
and according to research by market analysts Newzoo, in 2018, the global games market value across all platforms is around 135 billion USD.
The huge growth potential of the video game industry is also shown by the rapid sales increase during the coronavirus (COVID-19) pandemic.
According to the weekly sales data from GSD, 4.3 million games are sold globally during the week of March 16, 2020, which amounts to a rise of 63\% over the week prior.

\begin{figure}[htbp]
\begin{center}
\includegraphics[width=10cm]{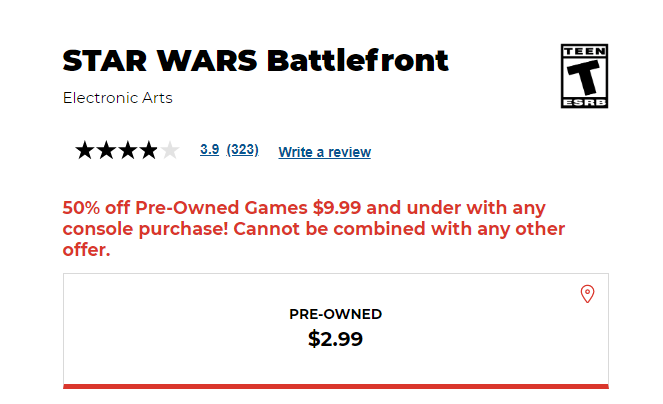}
\caption{Add-on discount example from Gamestop Corp, retrieved on 2019-08-18.}
\label{f_gamestop}
\end{center}
\end{figure}

Major platforms for video games include PCs, mobile phones, video game consoles and virtual reality (VR) headsets. Unlike PCs and mobile phones, video game consoles and VR headsets mainly support game functions. A unique structure for purchasing games for these devices is that customers have to first commit to the hardware, which is usually expensive, and then purchase the games, which are cheaper but include a large number of selections.
For retailers, this unique structure motivates a creative \emph{add-on discount strategy} for sales promotion, where a retailer offers customers discounts on a number of selected games after the customer makes a purchase of a video game console or a virtual reality headset.
Figure \ref{f_gamestop} shows an example of this strategy from Gamestop Corp., a major game retailer in the U.S. In this example, customers enjoy discounts on certain types of games if they purchase the games together with a game console.

\begin{figure}[htbp]
    \centering
    \subfloat[Combination 1.]{{\includegraphics[width=.8\linewidth]{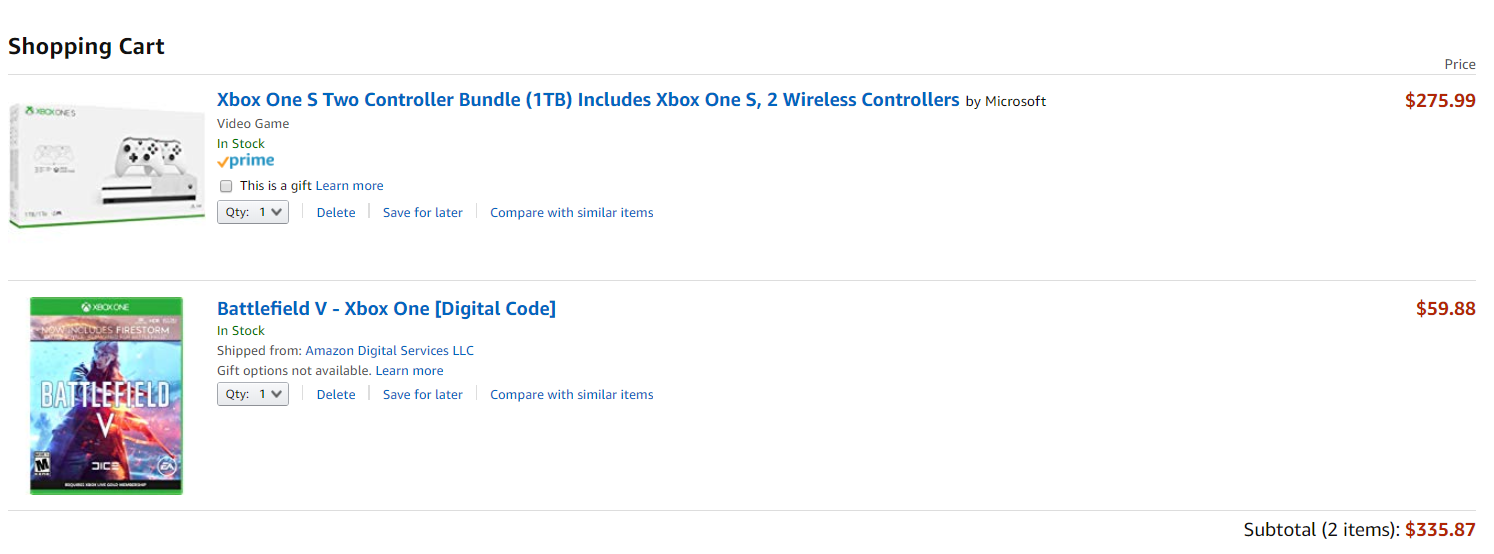} }}%
    \qquad
    \subfloat[Combination 2.]{{\includegraphics[width=.8\linewidth]{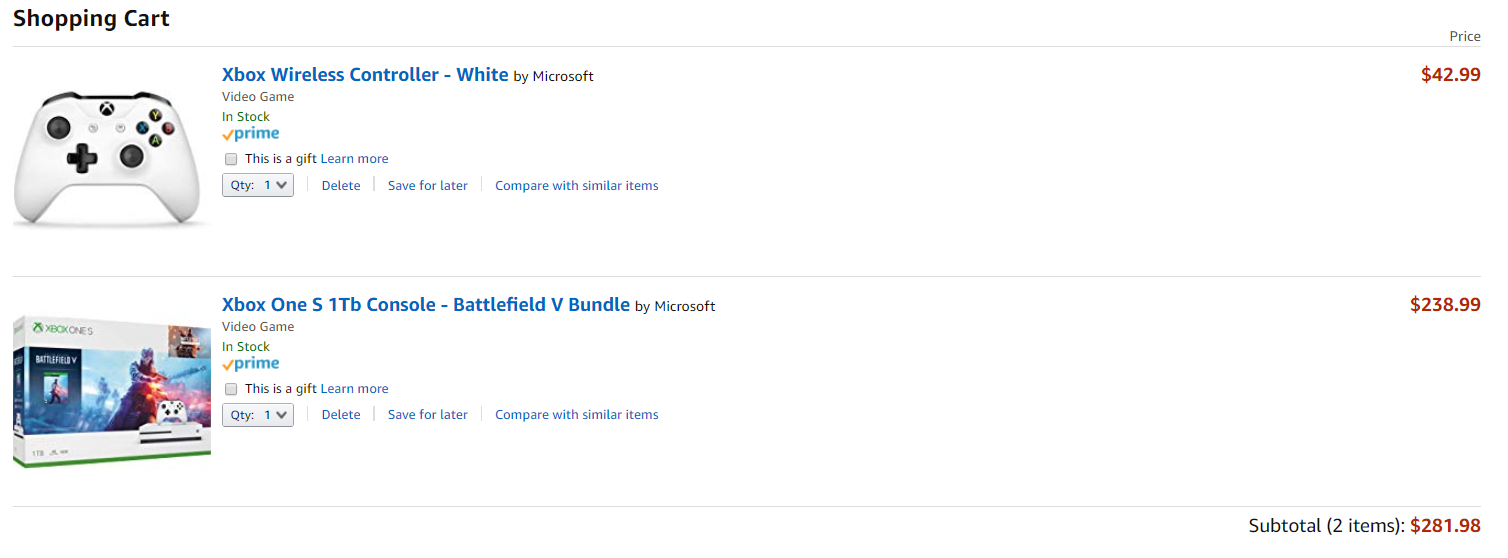} }}%
    \caption{Example of price inconsistency for the same selection of products.
    Retrieved on 2019-10-23 from Amazon.com.}%
    \label{f_2cases}%
\end{figure}

The add-on discount strategy is different from product bundling.
With add-on discounts, customers can make free selections from the offered set of add-on products, while with product bundling, customers can select only from fixed bundles of products. Moreover, although retailers can offer every possible product combination with add-on selection as a product bundle and decide each bundle's price individually,
such a strategy is not efficient in practice, and more importantly, might cause price inconsistencies.
Figure \ref{f_2cases} shows an example of price inconsistency.
In this example, consider a customer who wants to buy a game console, an extra controller and a certain game.
If the customer chooses combination 1, which contains a game and a console-controller bundle,
the final price would be $\$335.87$.
However, if the customer chooses combination 2, which contains a controller and a
console-game bundle, the final price would be $\$281.98$.
This significant price difference for the same selection of products results in an inconsistent environment, which not only creates bad shopping experience for customers, but also financially damages the retailer's business in the long run.
In contrast, for the add-on discount strategy, price inconsistency does not exist
because the final price always equals the sum of the prices of the selected products, and is thus independent from the way the products are combined.




The add-on discount structure exploits the complementary effects between products, and
thus adds new dimensions to the traditional pricing problem.
Specifically, in addition to deciding the original prices of different products,
retailers now also need to the decide the selection of add-on products, as well as their add-on discounts.
From the basic principle of optimization, we know that adding new dimensions
enlarges the feasible region of a problem, and hence leads to better decisions.
Therefore, by using the add-on discount strategy, a retailer can expect a higher revenue
than using the same pricing strategy with no add-on discounts.

Regardless of the advantages of the add-on discount strategy,
retailers may be hesitant to implement the strategy in practice
due to the following challenges.
One of the challenges is to limit the number of add-on discounts.
For example, when retailers show discount offers via pop-up messages on the customer's checkout page, there is usually a space limit on the total number of displayed offers.
In addition, if a retailer offers too many add-on discounts,
other retailers might take this as an arbitrage opportunity
by purchasing the products with discounts and selling them elsewhere at the original prices.
In these cases, retailers need to take \emph{space constraints} into account,
and the constraint increases the complexity of the problem.
Another challenge that might hold retailers back
is the lack of past experience or historical data.
In the scenario where retailers have no knowledge of the demand information,
blindly offering discounts with the add-on structure would harm the total revenue.
Hence retailers need to implement a learning algorithm together with the add-on discount strategy
to learn the unknown parameters on the fly,
and such design of the learning algorithm also increases the complexity of the problem.


\noindent \textbf{Our Contributions.}
In this paper, we study the revenue management problem with add-on discounts.
To the best of our knowledge, this is the first paper that formally studies this problem.
In particular, we consider a joint learning and optimization problem,
where the retailer does not know the demand functions of different products \emph{a priori},
and has to learn the information on the fly based on real-time observations of customers' purchases. Our formulation of the problem incorporates both the primary demand for products at their original prices and the add-on purchases for products with selected discounts.
We also consider a space limit constraint on the total number of add-on discount offers.

Our contributions in this paper can be summarized as follows.

\begin{itemize}

\item We formulate the revenue management problem with add-on discounts as an optimization problem with mixed binary decision variables.
In the offline setting where the retailer has access to all the demand information,
we develop an approximation algorithm that can solve the problem to any desired accuracy.
We also show that the algorithm is a Fully Polynomial-Time Approximation Scheme (FPTAS).

\item In the online setting where the retailer has no knowledge of the demand information, we develop an efficient UCB-based learning algorithm that uses FPTAS optimization algorithm as a subroutine. We show that the learning algorithm outputs a policy that converges to the offline optimal policy with a fast rate. We also show that the convergence rate is tight up to a certain logarithmic term.

\item We conduct numerical experiments based on the real-world transaction data we collect from Tmall.com. Based on our numerical results, we observe that our UCB-based learning algorithm has a robust performance and fast convergence rate in various test scenarios.
In addition, we observe that the learning algorithm can quickly outperform the optimal policy that does not use add-on discounts.
These observations illustrate the efficiency of our learning algorithm, as well as the
advantages of using the add-on discount strategy in practice.

\end{itemize}

\subsection{Literature Review}


To the best of our knowledge, the add-on discount strategy has not been formally studied in the the Operations Management (OM) literature, despite the fact that the strategy has been a common practice in the video game industry.
\cite{CMSX2019} study a similar but different model.
In that paper, the authors' focus is to figure out what products to recommend to a customer at the checkout stage, given the customer's primary purchase and each product's remaining inventory.
We highlight the difference between that paper and our paper as follows:
\begin{itemize}
\item In \cite{CMSX2019}, the focus is on the checkout stage, and they assume that customers' primary purchases are exogenous and not affected by the decision-maker. In contrast, in our model, the retailer controls both the primary purchase and the add-on purchase.
\item In their model, they assumed that when making the add-on recommendation for a certain products, there are two possible strategies: one is at the original price, and the other one is at a certain pre-determined discount price. In our model, we are not restricted to two alternatives.
\item In their model, they consider a fixed starting inventory. In our model, we do not include inventory.
\item From the methodology perspective, they focus on a competitive ratio analysis, and we consider the regret minimization.
\end{itemize}

Our work is also related to different areas of the literature: assortment planning, product bundling,  multi-armed bandit problems and UCB algorithms. Due to space limitation, we do not provide an exhaustive review of the literature and only provide a brief literature review as follows.

\textbf{Assortment Planning.}
The assortment planning problem models a customer's choice over a set of different products and focuses on finding the profit-maximizing assortment subject to various resource and capacity constraints. The problem has been studied extensively in the revenue management literature. In particular, in the offline setting where the underlying choice models are known, \cite{talluri2004revenue} propose an efficient algorithm for the single-resource assortment problem. \cite{gallego2004managing}, \cite{liu2008choice}, and \cite{zhang2009approximate} then extend the choice-based models to network revenue management problems.
Other works that study assortment algorithms under cardinality constraints, personalized decisions and various choice models can also be found in
\cite{kok2008assortment},
\cite{davis2013assortment}, \cite{golrezaei2014real},
\cite{cheung2016efficiency}, \cite{feldman2017revenue} and the references therein.

Recent research on assortment planning problems also focuses on the online setting where the parameters of the underlying choice models, such as multinomial logit (MNL), are not known and need to be learned online.
In this line of work, \cite{rusmevichientong2010dynamic}, \cite{agrawal2016near},
\cite{agrawal2017thompson}, \cite{agrawal2019mnl}, and \cite{MX2019}
study the problem where every customer follows the same choice model;
\cite{kallus2016dynamic}, \cite{cheung2017thompson}, \cite{bernstein2018dynamic}, \cite{miao2019context}, and \cite{miao2019fast} study the problem where each customer follows a personalized choice model.

Different from the assortment planning problems that mainly focus on how customers select one product from a set of alternatives, our model emphasizes customers' add-on purchase dynamics.

\textbf{Product Bundling.}
Both the add-on discount strategy and the bundling strategy are motivated by the complementary effects between products.
There exist various product bundling strategies in the literature,
such as pure bundling in which the retailer sells different products in a comprehensive bundle for a fixed price  (\cite{bakos1999bundling}),
mixed bundling in which the retailer offers all possible product bundles alongside individual products (\cite{chu2011bundle}), and customized bundling in which the retailer allows the customer to choose a certain quantity of products from a large pool of products for a fixed price (\cite{hitt2005bundling} and \cite{wu2008customized}).
We refer the readers to some recent papers (\cite{ma2015reaping}, \cite{abdallah2017large} and \cite{abdallah2019benefit}) for a more in-depth review of the bundling literature.

As mentioned in the example in Figure 2, add-on discounts and bundling are different.
The add-on discount strategy facilitates the customer's decision process, because with add-on discounts, the final price is only dependent on the set of products to purchase, not on how the bundles are formed.

\textbf{Multi-Armed Bandit (MAB) problems and UCB algorithms.}
The multi-armed bandit problem is a useful tool to study sequential decision-making problems under unknown rewards, and there exist a large number of papers studying this problem in the online learning literature. For a comprehensive review of the classic MAB algorithms and their performance analysis, see \cite{bubeck2012regret} and \cite{slivkins2019introduction}.

One of the classic multi-armed bandit models is the stochastic bandit, where the reward for pulling each arm is assumed to be i.i.d. drawn from an unknown probability distribution.
In the seminal paper \cite{auer2002nonstochastic}, the authors provide an algorithm that keeps updating the estimation of upper confidence bound (UCB) of each arm's mean reward, and show that such an algorithm can obtain an accumulative regret of $O(\sqrt{T\log T})$ in $T$ rounds.
The UCB-type algorithm is widely used in various bandit settings, such as linear bandits (\cite{rusmevichientong2010linearly}, \cite{abbasi2011improved}, \cite{chu2011contextual}),
combinatorial bandits (\cite{cesa2012combinatorial}, \cite{jin2019shrinking}),
and bandits with resource constraints (\cite{badanidiyuru2013bandits}, \cite{agrawal2016linear}).

In the OM literature, recent research papers have also been focusing on problems under uncertain environments and applying bandit algorithms or other learning algorithms to tackle the exploration-exploitation tradeoffs in learning tasks.
This includes dynamic pricing problems (\cite{besbes2009dynamic}, \cite{besbes2012blind}, \cite{wang2014close}, \cite{BZ2015},
\cite{ferreira2018online}, \cite{gao2018multi})
and inventory control problems with unknown demand distributions
(\cite{zhang2017perishable}, \cite{zhang2019closing},
\cite{chen2019coordinating}, \cite{YLS2019}),
assortment optimization problems with unknown purchase probabilities
(\cite{cheung2017assortment}, \cite{agrawal2019mnl}),
online matching and resource allocation problems with unknown reward distributions (\cite{cheung2018inventory}).

\noindent \textbf{Organization of the paper.}
The remainder of the paper is organized as follows.
In Section \ref{sec_mod}, we present the formulation of the revenue management problem with add-on discounts.
Then, in Section \ref{sec_opt}, we study the offline optimization problem and propose an approximation algorithm that can solve the problem to any desired accuracy.
In Section \ref{sec_learn}, we consider the online setting where the demand functions of different products are unknown.
We present algorithm $\ucb$, a UCB-based algorithm, to solve the online problem, and show the performance of the algorithm through regret analysis.
Next, in Section \ref{sec_dis}, we discuss several model assumptions, possible extensions and variants of our formulation.
In Section \ref{sec_num}, we present the results of our numerical experiments which are based on the real-world transaction data we collect from Tmall.com.
We conclude the paper with a discussion of future research directions in Section \ref{sec_con}.

%% file: Model.tex
\section{Model} \label{sec_mod}

We present in this section the formulation of the revenue management problem with add-on discounts.

Consider a retailer managing two types of products: core products (e.g., different variants of a video game console from the same brand) and supportive products (e.g., video games for the same brand of video game consoles).
Let $N$ denote the number of core products, indexed by $\{1,\ldots, N\}$,
and $M$ the number of supportive products, indexed by $\{1,\ldots, N\}$.
For each core product $n=1,\ldots, N$, we assume that its price $p_n$ is selected from set $\Omega_c:=\{q_c^{1},q_c^{2},...,q_c^{\zeta(c)}\}$, and
for each supportive product, $m=1,\ldots, M$, we assume that its price $p_{N+m}$ is selected from set $\Omega_s:=\{q_s^{1},q_s^{2},...,q_s^{\zeta(s)}\}$.
Let the binary variable $I_{N+m}\in\{0,1\}$ with $m=1,\ldots, M$ denote whether or not we offer an add-on discount for supportive product $m$.
Denote the add-on discount price for product $m$ as $p_{N+m}'$, and we assume
$p_{N+m}'$ is selected from $\Omega_{a}:=\{q_{a}^{1},q_{a}^{2},...,q_{a}^{\zeta(a)}\}$.
In addition, as we discuss in Section \ref{sec_intro}, the retailer cannot offer too many add-on discounts. Thus, we consider in our model an additional space constraint that limits the total number of add-on discounts to be within $S$, i.e.,
$\sum_{m=1}^{M} I_{N+m} \leq S$.

\begin{figure}[htbp]
\begin{center}
\includegraphics[width=13cm]{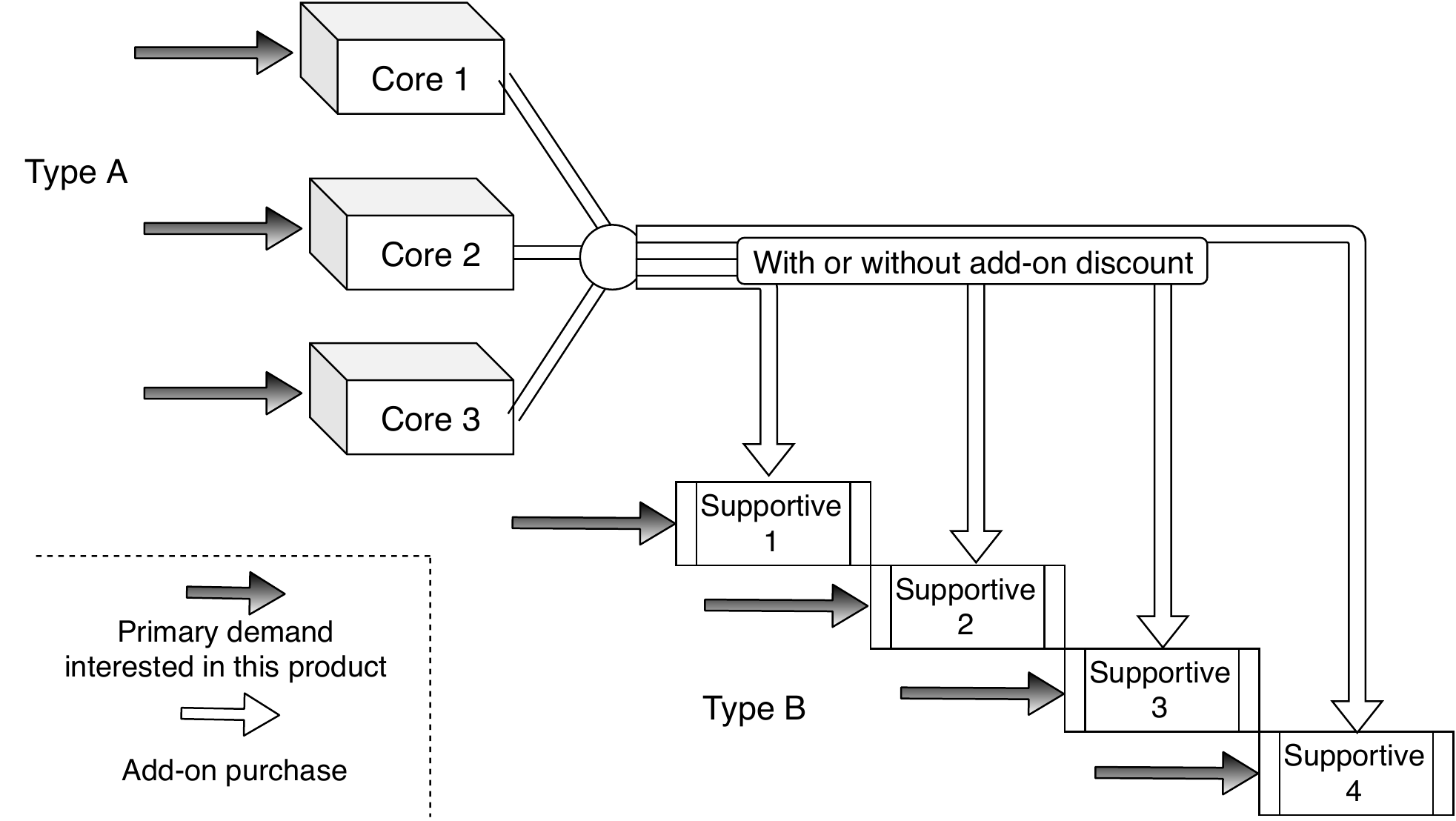}
\caption{Illustration of a customer's purchase dynamics.}
\label{f_dynamics}
\end{center}
\end{figure}

On the demand side, there exist two types of purchases, differentiated by whether or not a customer has already owned a core product before they arrive at the retailer's online store.
Since we consider the core product as a prerequisite for using supportive products (e.g., video game console for video games), we assume that customers will not consider purchasing any supportive product without owning or purchasing a core product first.
This condition results in two types of purchases: A) purchases from customers that do not own a core product,  and B) purchases from customers that have already owned a core product.
For type A purchases, customers will first purchase a core product, and then they may or may not continue to purchase supportive products with or without add-on discounts.
For type B purchases, customers will only consider purchasing supportive products without any add-on discounts.
The purchase dynamics are further illustrated in Figure \ref{f_dynamics}.
In addition, we partition the purchases into two categories: primary demand and add-on purchase, as indicated by the colors of the arrows in the figure.

The entire selling horizon is divided into discrete time periods.
We assume that each time period is short enough so that the primary demand for each core and each supportive product is a Bernoulli random variable.
In particular, we use $\alpha_n(p_n)$ to denote the primary demand for core product $n=1,\ldots,N$, and $\alpha_{N+m}(p_{N+m})$ the primary demand for supportive product $m=1,\ldots,M$.

The add-on purchase category involves purchases both with and without discounts, depending on if we are offering add-on discount for each product, and we differentiate them as follows.
\begin{itemize}
\item  Let $\beta_{N+m}'(p_{N+m}')$ be the probability that a customer continues to purchase product $N+m$ under discount price $p_{N+m}'$, after she purchases one core product.
\item Let $\beta_{N+m}(p_{N+m})$ be the probability that a customer continues to purchase product $N+m$ under original price $p_{N+m}$ , after she purchases one core product.
\end{itemize}

We also assume that all demand parameters are independent across different products.
We will discuss these model assumptions in detail in Section \ref{sec_dis}.

Let $\mathcal{R}$ denote the expected revenue per time period (each time period is identical). Given the retailer's goal of maximizing the total expected revenue, we can formulate the revenue management problem as:

\begin{equation}
\label{opt_problem}
\begin{aligned}
    \max  _{p_n, p_{N+m}, p_{N+m}', I_{N+m}}
    \quad & \mathcal{R} := \sum_{n=1}^{N} \alpha_n(p_n)p_n + \sum_{m=1}^{M}\alpha_{N+m}(p_{N+m})p_{N+m}  \\
    & \quad \quad + \left [ \sum_{n=1}^{N}\alpha_n(p_n) \right ] \cdot \sum_{m=1}^{M} \left [I_{N+m} \cdot \beta_{N+m}'(p_{N+m}^{'}) \cdot p_{N+m}^{'} \right ]  \\
    & \quad \quad + \left [ \sum_{n=1}^{N}\alpha_n(p_n) \right ] \cdot \sum_{m=1}^{M} \left [ \left ( 1- I_{N+m} \right )\cdot \beta_{N+m}(p_{N+m}) \cdot p_{N+m} \right ]\\
    \textrm{s.t.} \quad & \sum_{m=1}^{M} I_{N+m} \le S ,  \\
    & p_{N+m}' < p_{N+m} \text{ for } m=1,\ldots, M,  \\
    & p_n\in \Omega_c \text{ for } n=1,\ldots, N,\\
    & p_{N+m} \in \Omega_s,  \;
    p_{N+m}'\in \Omega_{a},  \; I_{N+m}\in \{0,1\} \text{ for } m=1,\ldots, M.
\end{aligned}
\end{equation}


In this optimization problem, the set of decisions include:
the original price for each core product,
the original and add-on discount price for each supportive product,
and the binary indicator on whether or not to select each supportive product for add-on discount.
The first term in $\mathcal{R}$ corresponds to the primary demand for core products, and the second term corresponds to the primary demand for supportive products. The third and fourth terms correspond to the add-on purchases for supportive products with and without add-on discounts, respectively.
The first constraint sets the space constraint (upper bound) on the total number of add-on discounts.
The second constraint requires that the discount price is less than the original price.

We observe from the formulation that the optimization problem is difficult to solve because
the problem contains 1) discrete decision variables and 2) products of decision variables.
In addition, the total number of feasible solutions is exponentially large, which makes the enumeration method intractable. Therefore, instead of finding the exact optimal solution, we propose in this paper an approximation algorithm that can solve the problem to any desired accuracy. We also show that the algorithm is an FPTAS, which means the running time of the algorithm is polynomial in both the problem size and the approximation error.

The optimization problem provides solutions to the revenue management problem in the offline setting where the demand functions $\alpha_n(\cdot)$, $\alpha_{N+m}(\cdot)$, $\beta_{N+m}(\cdot)$ and $\beta_{N+m}'(\cdot)$ are known.
However, in practice, this information may not be available to the retailer due to the lack of historical transaction data, and the retailer then needs to learn the parameters online.
In the following sections, we first present our solution to the offline optimization problem in Section \ref{sec_opt}. Then in Section \ref{sec_learn}, we propose a UCB-based learning algorithm that uses the offline optimization algorithm as a subroutine to solve the problem in the online setting.

%% file: Optimization.tex
\section{Optimization Subroutine} \label{sec_opt}

In this section, we propose an approximation algorithm that
can solve the offline optimization problem \eqref{opt_problem}
to any desired accuracy, and show that the algorithm is an FPTAS.

As discussed in Section \ref{sec_mod}, the optimization problem is challenging due to the existence of discrete decision variables and products of decision variables.
To resolve these challenges, we reformulate the original problem into two parts that separate the purchase of core products and the purchase of supportive products.
We refer to these two problems as the \emph{master} problem and the \emph{subproblem}, respectively.

In the decomposed formulation, we replace the term $\sum_{n=1}^{N}\alpha_n(p_n)$ with $\gamma$, which represents the demand of core products per period. In addition, we introduce function $\mathcal{R}_s(\gamma)$ to denote the optimal revenue from the purchase of supportive products, which includes primary demand $\alpha_{N+m}(\cdot)$, add-on purchase $\beta_{N+m}(\cdot)$ and $\beta_{N+m}'(\cdot)$, when the demand for the core products is $\gamma$.

\begin{equation}
\label{eq:master}
\begin{aligned}
    \textbf{Master problem.} \quad \max_{p_n} \quad & \mathcal{R}
    =\sum_{n=1}^{N}\alpha_n(p_n)p_n + \mathcal{R}_s(\gamma)\\
    \textrm{s.t.} \quad  & \sum_{n=1}^{N}\alpha_n(p_n)= \gamma \\
    & p_{n} \in \Omega_c.
\end{aligned}
\end{equation}

\begin{align} \label{eq:subproblem}
    \textbf{Subproblem.} \quad  \mathcal{R}_s(\gamma) := \max _{p_{N+m}, p'_{N+m}, I_{N+m}} \quad & \sum_{m=1}^{M}\alpha_{N+m}(p_{N+m})p_{N+m} \nonumber \\
    & \quad \quad +  \gamma  \cdot \sum_{m=1}^{M} \left [I_{N+m} \cdot \beta_{N+m}'(p_{N+m}^{'}) \cdot p_{N+m}^{'} \right ] \nonumber \\
    & \quad \quad + \gamma  \cdot \sum_{m=1}^{M} \left [ \left ( 1- I_{N+m} \right )\cdot \beta_{N+m}(p_{N+m}) \cdot p_{N+m} \right ] \\
    \textrm{s.t.} \quad & \sum_{m=1}^{M} I_{N+m} \le S , \nonumber \\
    & p_{N+m} < p_{N+m}' \text{ for } m=1,\ldots, M   \nonumber  \\
    & p_{N+m}\in \Omega_{s}, \; p_{N+m}'\in \Omega_{a},  \; I_{N+m}\in \{0,1\} \text{ for } m=1,\ldots, M. \nonumber
\end{align}

The decomposed formulation does not provide a tractable solution directly: in order to solve the problem, we need to determine the value of $\gamma$, which can take exponentially many values within $[0,N]$.
Nevertheless, since $\gamma$ is bounded, we can adopt a discretization approach that solves the problem for only a set of discrete points in $[0,N]$.
In the following, we first show in Lemma \ref{lem_Lip} that function $\mathcal{R}_s(\gamma)$ is Lipschitz continuous.
Then building on this lemma, we develop an approximation algorithm using the discretization approach to solve the master problem.

The high-level intuition for function $\mathcal{R}_s(\gamma)$'s Lipschitz continuity is based on the observation that parameter $\gamma$ appears in the objective function of the subproblem. Thus, when the value of $\gamma$ changes locally, the value of $\mathcal{R}_s(\gamma)$ should not change too much.

\begin{lemma} \label{lem_Lip}
The function $\mathcal{R}_s(\gamma)$, as defined in \eqref{eq:subproblem}, is Lipschitz continuous in $\gamma \geq 0$ with parameter $M\cdot \hat{p}$, where $\hat{p}$ is the highest price among all the products, namely,
$\hat{p}:= \max_{p \in \Omega_c \cup \Omega_s} p$.
\end{lemma}

\textit{Proof.}
Note that $\mathcal{R}_s(\gamma)$ represents the optimal revenue from supportive products,
given that the expected total sales from core products are $\gamma$.


By definition \eqref{eq:subproblem}, we can reformulate $\mathcal{R}_s(\gamma)$ as
\begin{equation}\label{eq:subproblem_reformulation}
\begin{aligned}
\mathcal{R}_s(\gamma) = \min _\pi & \quad x \\
\textrm{s.t.} \quad & x \geq \gamma \cdot F_2(\pi) + F_1(\pi) ,
    \;\forall \; \text { feasible policy } \pi ,
\end{aligned}
\end{equation}
where $\pi$ denotes the feasible policy of the subproblem.

Formally, the feasible policy $\pi$ is defined by the feasible solution to problem \eqref{eq:subproblem}, which specifies the values $p_{N+m} \in \Omega_s$, $p'_{N+m}\in \Omega_{sa}$ and $I_{N+m} \in \{0,1\}$, that satisfy $p_{N+m}>p'_{N+m}, \forall\; m=1,\ldots,M$ and $\sum_{m=1}^{M} I_{N+m}\leq S$.
Function $F_1(\pi)$ and $F_2(\pi)$ are defined as
\begin{eqnarray*}
F_1(\pi) &:=&  \sum_{m=1}^{M} \alpha_{N+m}(p_{N+m})p_{N+m} \\
F_2(\pi) &:=&  \sum_{m=1}^{M} I_{N+m} \beta_{N+m}' (p_{N+m}')p_{N+m}' +
    (1-I_{N+m}) \beta_{N+m}(p_{N+m})p_{N+m}.
\end{eqnarray*}

Observe that in this reformulation, $F_1(\pi)$ and $F_2(\pi)$ are constants, and
since the number of feasible policies is finite, the total number of constraints in \eqref{eq:subproblem_reformulation} is also finite.
Moreover, the RHS of each constraint is a linearly increasing function of $\gamma$.
Hence we know that for any $\gamma$, the optimal solution $x$ is equal to
\begin{equation*}
\max_{\pi} \;\{\gamma \cdot F_2(\pi) + F_1(\pi)\},
\end{equation*}
and we obtain
\begin{equation*}
\mathcal{R}_s(\gamma) = \min\; \max_{\pi}\; \{\gamma \cdot F_2(\pi) + F_1(\pi)\}.
\end{equation*}

Therefore, we know that $\mathcal{R}_s(\gamma)$ is a convex piece-wise linear function,
and it implies that $\mathcal{R}_s(\gamma)$ is Lipschitz continuous.
Specifically, the Lipschitz parameter is equal to the function's maximum \emph{slope}, i.e.,
$\max_{\pi} F_2(\pi)$, which is bounded by $M\cdot \hat{p}$, by definition. \Halmos

%

Lemma \ref{lem_Lip} implies that we can approximate the value of $\mathcal{R}_s(\gamma)$  with a guaranteed accuracy. More importantly, this result motivates an approximation scheme
where we only need to evaluate the value of $\mathcal{R}_s(\gamma)$ for a set of discrete points in $[0,N]$, instead of for all possible $\gamma$ values.

Based on the approximation scheme, we can develop the solutions to the subproblem and the master problem separately.
Specifically, for the subproblem, we can formulate it as a selection problem and solve it using a greedy approach.
For the master problem, we can formulate it as a $N$-stage dynamic program, with approximations between stages, and solve it using backward induction.

We formally describe the detailed procedures of our algorithm in Algorithm \ref{Alg_sub}.
Then we show in Lemma \ref{lem_run_time} that Algorithm \ref{Alg_sub} has a polynomial runtime. Next, in Lemma \ref{lem_gap}, we show that Algorithm \ref{Alg_sub} has a bounded approximation error. Building on the results of these two lemmas, we show in Theorem \ref{thm_FPTAS} that Algorithm \ref{Alg_sub} is an FPTAS.


\begin{algorithm}[!]
\caption{FPTAS optimization subroutine for the offline optimization problem:}
\label{Alg_sub}
\begin{itemize}
\item \textbf{Algorithm input:}
\begin{itemize}
\item $\Omega_c$,  $\Omega_s$, $\Omega_{a}$,
\item $\alpha_n(p_n)$ for all $p_n\in \Omega_c$ and $n=1,\ldots, N$,
\item $\alpha_{N+m}(p_{N+m})$ for all $p_{N+m}\in \Omega_s$ and $m=1,\ldots, M$,
\item $\beta_{N+m}'(p_{N+m}')$ for all $p_{N+m}' \in \Omega_{a}$ and $m=1,\ldots, M$,
\item $\beta_{N+m}(p_{N+m})$ for all $p_{N+m}\in \Omega_{s}$ and $m=1,\ldots, M$,
\item Integer constant $K$.
\end{itemize}

\item \textbf{Part 1:} Solve supportive revenue part separately.
\begin{enumerate}[label=\alph*)]
\item For all $m=1,\ldots, M$, and $\gamma=0,\frac{1}{K},\frac{2}{K},\ldots,\frac{NK}{K}$, solve
$$\max_{p_{N+m}\in \Omega_s } \alpha_{N+m}(p_{N+m})p_{N+m}+\gamma \beta_{N+m}(p_{N+m})p_{N+m},$$
and denote the optimal objective value as $r_{N+m}(\gamma)$.
\item For $m=1,\ldots, M$, and $\gamma=0,\frac{1}{K},\frac{2}{K},\ldots,\frac{NK}{K}$, solve
$$\max_{p_{N+m}\in \Omega_s, p_{N+m}'\in \Omega_{a}, p_{N+m}' < p_{N+m} } \alpha_{N+m}(p_{N+m})p_{N+m}+ \gamma \beta_{N+m}'(p_{N+m}')p_{N+m}',$$
and denote the optimal objective value as $r_{N+m}'(\gamma)$.
\item For $\gamma=0,1/K,2/K,\ldots,N$, sort the values of
$r_{N+m}'(\gamma)-r_{N+m}(\gamma)$ into an array
in the descending order.
Set $I_{N+m}(\gamma)=1$, if $r_{N+m}'(\gamma)-r_{N+m}(\gamma)$ is positive and in the first $S$ entries of the array of sorted values. Set $I_{N+m}(\gamma)=0$, otherwise.

\item For $\gamma=0,1/K,2/K,\ldots,N$, let
$$\mathcal{R}_s(\gamma) = \sum_{m=1}^{M} \left[r_{N+m} +
I_{N+m}(\gamma) \cdot [r_{N+m}'(\gamma)-r_{N+m}(\gamma)] \right].$$
\end{enumerate}
\vspace{-10pt}
\item \textbf{Part 2:} Combining the revenue of core products and supportive products using dynamic programming.

\begin{enumerate}[label=\alph*)]
\item \textit{Initialization:} For $n=1,\ldots,N$ and $p_n \in \Omega_c$, let $\hat{\alpha}_n(p_n)$ be $\alpha_n(p_n)$ rounded to the nearest integer multiple of $1/K$.
\item \textit{State:} $(n,\gamma)$. \textit{Action:} $p_n$ in every state $(n,\boldsymbol{\cdot})$.
\item \textit{Value function:} $V_n(\gamma)$ is defined as the maximum revenue to be earned from all the products (both core and supportive) excluding product $1$ to $n-1$, when the total (approximate) demand for the first $n-1$ products are $\gamma$.
\item \textit{Optimality equation:} $V_n(\gamma)=\max_{p_n\in \Omega_c} \left [ \alpha_n(p_n)\cdot p_n + V_{n+1}(\gamma + \hat{\alpha}_n(p_n)) \right ] .$
\item \textit{Boundary condition:} $V_{N+1}(\gamma)= \mathcal{R}_s(\gamma) $, for all $\gamma=0,1/K,2/K,\ldots,N$.
\item The above DP can be solved efficiently using backward induction, the optimal decisions can be retrieved along the optimality equations, and $V_1(0)$ is the approximate optimal total revenue.
\end{enumerate}

\end{itemize}
\end{algorithm}

\begin{lemma} \label{lem_run_time}
Algorithm \ref{Alg_sub} has a runtime of complexity $O(C^4\cdot K)$, where
$$C:=\max(M,N,\left| \Omega_c \right|, \left| \Omega_s \right|, \left| \Omega_{a} \right|).$$
\end{lemma}

\textit{Proof.} Consider the algorithm's runtime in the Big-O complexity.
In Part 1 of Algorithm \ref{Alg_sub}, step b) takes the longest runtime.
Specifically, in step b), we enumerate $M\cdot N\cdot K$ cases in total,
and solve each case by enumerating all possible pairs of $p_{N+m}$ and $p'_{N+m}$ such that $p_{N+m}> p'_{N+m}$. The runtime for step b) is thus
$O\left(M\cdot N\cdot K\cdot \left| \Omega_s \right| \cdot \left| \Omega_{a} \right|\right)\leq O\left( C^4\cdot K \right)$,
and this also gives the runtime complexity of Part 1.
In Part 2 of Algorithm \ref{Alg_sub}, the total number of states is
$O\left( C^2\cdot K \right)$.
In addition, for each state, we check the optimality equation once,
which has runtime $O\left( C \right)$.
The runtime complexity for Part 2 is thus $O\left( C^3\cdot K \right)$.
Combining the two parts, we obtain the algorithm's total runtime complexity
$O\left ( C^4\cdot K \right)$.
\Halmos

\begin{lemma} \label{lem_gap}
The approximation error of Algorithm \ref{Alg_sub} is upper-bounded by
$$\frac{\hat{p}MN}{K},$$ where $\hat{p}$ is the highest price among all the products.
\end{lemma}

\textit{Proof.} Let $V(\pi)$ be the \emph{true} revenue of policy $\pi$, and $V'(\pi)$ the \emph{approximate} revenue of policy $\pi$ that is provided by Algorithm \ref{Alg_sub}.
In addition, let OPT be the optimal policy of problem \eqref{opt_problem}, and
ALG the ``optimal" policy that is provided by Algorithm \ref{Alg_sub}.

Given policy $\pi$, we know that $V(\pi)$ and $V'(\pi)$ give the same revenue for the core products, but different revenue for the supportive products.
Specifically, due to the rounding procedure, the value of $\gamma$ we use in Algorithm \ref{Alg_sub} differs from its \emph{true} value by at most $N/2K$.
Hence by Lemma \ref{lem_Lip}, we have
\begin{equation} \label{eq_gap}
\left | V(\pi) - V'(\pi)  \right |\le \frac{\hat{p}MN}{2K},
\end{equation}
for any feasible policy $\pi$.

Therefore, we have
\begin{align*}
V(OPT)&\leq V^{'}(OPT)+\frac{ \hat{p}MN}{2K}\\
&\leq V^{'}(ALG)+\frac{\hat{p}MN}{2K}\\
&\leq  V(ALG)+\frac{\hat{p}MN}{K},
\end{align*}
where the first and last inequality follow \eqref{eq_gap}.
The second inequality follows because ALG optimizes the approximate revenue $V'(\cdot)$. \Halmos




\begin{theorem} \label{thm_FPTAS}
Suppose $V(OPT)\ge v^*$. For any problem instance and an $\varepsilon>0$,
Algorithm \ref{Alg_sub} can output an $(1-\varepsilon)$-optimal policy, with running time polynomial in both the problem size and $1/\varepsilon$, with parameter
$$K=\left \lceil \frac{\hat{p}MN}{v^* \cdot \varepsilon } \right \rceil.$$
In other words, Algorithm \ref{Alg_sub} is an FPTAS.
\end{theorem}

\textit{Proof.}
By Lemma \ref{lem_gap}, we know that the approximation error of Algorithm \ref{Alg_sub} is bounded by $\frac{\hat{p}MN}{K}$. Given the value of $K$, we have
\begin{equation*}
    V(OPT) - V(ALG) \leq \frac{\hat{p}MN}{K} \leq \cdot \varepsilon v^*.
    \leq \; \varepsilon \cdot V(OPT).
\end{equation*}
Therefore, the algorithm is $(1-\varepsilon)$-optimal.
By Lemma \ref{lem_run_time}, we also know that the algorithm's runtime is polynomial in both the problem size and $1/\varepsilon$. Therefore, Algorithm \ref{Alg_sub} is an FPTAS. \Halmos

%% file: Learning.tex
\section{Learning Algorithm and Regret Analysis} \label{sec_learn}

We consider in this section the revenue management problem in the online setting where
the demand functions $\alpha_n(\cdot)$, $\alpha_{N+m}(\cdot)$, $\beta_{N+m}(\cdot)$ and $\beta_{N+m}'(\cdot)$ are not known \emph{a priori}.
In this setting, the retailer needs to determine the prices of different products and the selection of products with add-on discounts, while conducting price experiments and learning the demand information on the fly.
More importantly, with the goal of maximizing the total revenue over $T$ selling periods,
the retailer faces the classic learning (exploration) and earning (exploitation) trade-off.


To tackle these challenges from unknown demand parameters,
we model the joint learning and optimization problem as a multi-armed bandit,
and develop a UCB-based algorithm to solve the problem.
One way to design the algorithm is to construct the upper confidence bound (UCB) of the expected revenue (i.e., reward) of each policy (i.e., arm), which is equal to the empirical mean of each policy's revenue plus a confidence interval.
Then the algorithm picks the policy with the highest upper confidence bound in each period.
However, this naive construction of the UCBs results in the following issues.

\begin{itemize}
    \item The learning algorithm is highly inefficient
    because the total number of policies in our problem is exponentially large.
    Consequently, the regret of this learning algorithm, as defined in \eqref{eq:def_regret}, would be very large, meaning the algorithm can hardly converge to the optimal policy.
    \item In each period of the algorithm, it is impossible to compare an exponential number of policies so as to find the best one to implement.  In addition, it is difficult to implement the learning algorithm together with the optimization subroutine we propose in Section \ref{sec_opt}.
\end{itemize}



To resolve these issues, we adopt an alternative way of constructing the UCBs:
instead of estimating the UCBs for each policy,
we estimate the UCBs for each unknown parameter, namely, $\alpha_n(p_n)$,
$\alpha_{N+m}(p_{N+m})$, $\beta_{N+m}'(p_{N+m}')$ and $\beta_{N+m}(p_{N+m})$,
for $p_n\in \Omega_{c}$, $p_{N+m} \in \Omega_{s}$ and $p_{N+m}' \in \Omega_{a}$.
Then, we can use these estimates as inputs to the FPTAS optimization subroutine to
determine the ``optimal" policy in each period.

\subsection{The learning algorithm}

In the UCB-based learning algorithm, we keep track of the empirical mean of demand parameters
$\alpha_n(p_n)$, $\alpha_{N+m}(p_{N+m})$, $\beta_{N+m}'(p_{N+m}')$, $\beta_{N+m}$
for all products $n\in\{1,\ldots,N\}$, $m\in\{1,\ldots,M\}$ and all prices
$p_n\in \Omega_{c}$, $p_{N+m} \in \Omega_{s}$, $p_{N+m}' \in \Omega_{a}$, respectively.
We also keep track of the \emph{counter} of each price, which counts
the number of periods or the number of purchased products associated with the price,
for each type of demand function.

We introduce the following notations in our algorithm.


\begin{itemize}

\item $\overline{\alpha}_n(p_n)$: the empirical average of $\alpha_n(p_n)$, for all $n=1,...,N$ and $p_n\in \Omega_c$.
\item $\overline{\alpha}_{N+m}(p_{N+m})$: the empirical average of $\alpha_{N+m}(p_{N+m})$, for all $m=1,...,M$ and $p_{N+m}\in \Omega_s$.
\item $\overline{\beta}_{N+m}'(p_{N+m}')$: the empirical average of $\beta_{N+m}'(p_{N+m}')$, for all $m=1,...,M$ and $p_{N+m}'\in \Omega_{a}$.
\item $\overline{\beta}_{N+m}(p_{N+m})$: the empirical average of $\beta_{N+m}(p_{N+m})$, for all $m=1,...,M$ and $p_{N+m}\in \Omega_s$.
\item $c_n(p_n)$: the number of periods that price $p_n$ of core product $n$ has been used, for all $n=1,...,N$ and $p_n\in \Omega_c$.
\item $c_{N+m}(p_{N+m})$: the number of periods that price $p_m$ of supportive product $N+m$ has been used, for all $m=1,...,M$ and $p_{N+m}\in \Omega_s$.
\item $c_{N+m}^{(a,1)}(p_{N+m}')$: the number of core products purchased when product $N+m$ is selected as an add-on product under the discount price $p_{N+m}'$, for all $m=1,...,M$ and $p_{N+m}' \in \Omega_{a}$.
\item $c_{N+m}^{(a,2)}(p_{N+m})$: the number of core products purchased when
product $N+m$ is not selected as an add-on product but offered at the original price $p_{N+m}$, for all $m=1,...,M$ and $p_{N+m} \in \Omega_{s}$.
\end{itemize}


We also introduce the notion of \emph{episode}, which is defined as a consecutive number of periods.
In the algorithm, we update the ``online" policy at the beginning of each episode, and then use the policy for a number of periods, until the episode terminates with certain \emph{stopping rules}.
Therefore, the length of each episode (in periods) is in fact a stopping time.

We refer to our learning algorithm as $\ucb$, and formally describe it in Algorithm \ref{Alg_UCB}.

In the beginning of each episode, the algorithm first uses the FPTAS optimization subroutine to solve an \emph{optimistic} version of the problem, in which all the parameters are evaluated at their UCBs. In addition, given that the demand parameters are defined as Bernoulli random variables, we truncate all the UCBs at value $1$.

We also observe that the parameter $K$ increases as the number of episodes $\tau$ and time period $t$ increase. By Theorem \ref{thm_FPTAS}, we know the approximation error of the optimization subroutine decreases with $\tau$, while the computation time increases.
To mitigate the computational cost, we set the stopping criteria where
given policy $\Pi_{\tau}$, each episode ends when
the value of at least one of the associated counters is doubled within the episode.
By this construction, the length of each episode increases in $\tau$.
As a result, the algorithm calls the subroutine less frequently as time increases, and the output policy also becomes more stable.

\begin{algorithm}[t]
\caption{$\ucb$}
\label{Alg_UCB}
\begin{itemize}

\item \textbf{Initialization.}

Set all the empirical means and counters to $0$.
Set the value of $\varepsilon$.

\item \textbf{Loop. For each episode $\tau$,}

\begin{itemize}

\item[1.] Let period $t$ denote the first period in $\tau$. Solve the optimization problem \eqref{opt_problem} using the FPTAS subroutine with the following inputs.

\begin{itemize}
\item $\Omega_c$,  $\Omega_s$, $\Omega_{a}$,
\item $\widetilde{\alpha}_n(p_n)
    =\min\left (1,\bar{\alpha}_n(p_n)
    + \sqrt{\frac{2\log t}{c_n(p_n)}}\right )$
    for all $p_n\in \Omega_c$ and $n=1,\ldots, N$
\item $\widetilde{\alpha}_{N+m}(p_{N+m}) :=
    \min \left (1,\bar{\alpha}_n(p_{N+m})
    + \sqrt{\frac{2 \log t}{c_{N+m}(p_{N+m})}} \right )$
    for all $p_{N+m}\in \Omega_s$ and $m=1,\ldots, M$
\item $\widetilde{\beta}_{N+m}'(p_{N+m}') :=
    \min\left (1,\bar{\beta}_{N+m}'(p_{N+m}')
    + \sqrt{\frac{2\log t}{c^{(a,1)}_{N+m}(p_{N+m}')}}\right )$
    for all $p'_{N+m}\in \Omega_{a}$ and $m=1,\ldots, M$
\item $\widetilde{\beta}_{N+m}(p_{N+m})=
    \min\left (1,\bar{\beta}_{N+m}(p_{N+m})
    + \sqrt{\frac{2 \log t}{c^{(a,2)}_{N+m}(p_{N+m})}} \right) $
    for all $p_{N+m}\in \Omega_{s}$ and $m=1,\ldots, M$

\item $K= \left \lceil \frac{\sqrt{t}}{\varepsilon } \right \rceil$
\end{itemize}

\item[2.] Denote the output policy as $\Pi_{\tau}$. Keep using policy $\Pi_{\tau}$ and updating the empirical means and counters in each period, accordingly.

\item[3.] Terminate the episode when the value of at least one of the counters (associated with the selected add-on products and prices under policy $\Pi_{\tau}$)
is doubled within the episode.

\item[] Set $\tau=\tau+1$.

\end{itemize}
\end{itemize}
\end{algorithm}

\subsection{Regret analysis}


We analyze the performance of our learning algorithm by adopting the standard notion of regret.
Let $\mathcal{R}^*$ be the one-period expected revenue of the optimal \emph{clairvoyant} policy that has access to the full demand information,
and $\mathcal{R}(\Pi_t)$ the expected revenue of the policy $\Pi_t$ that is used by algorithm $\ucb$ in period $t$.
The regret of our algorithm is then defined as

\begin{equation}\label{eq:def_regret}
\text{\emph{Regret}}(T) := \mathbf{E} \left [\sum_{t=1}^T  \mathcal{R}^* - \mathcal{R}(\Pi_t) \right ].
\end{equation}
Since $\mathcal{R}^*$ is an upper bound of $\mathcal{R}(\Pi)$ for any policy $\Pi$,
the regret is always non-negative.
In the following theorem, we state our main result on
the upper bound of our learning algorithm's regret.


\begin{theorem}
\label{main}
For any problem instance, the regret of algorithm $\ucb$ can be upper-bounded by
\begin{equation}\label{eq:thm_regret}
\text{Regret}\;(T) \leq \mathcal{O} \left( N M\hat{p}
    \left( (1/\lambda) \cdot \sqrt{U T\log T} +
    \varepsilon \sqrt{T}  \right) \right),
\end{equation}
where $\hat{p}$ is the maximum price as defined in Lemma \ref{lem_Lip},
$\lambda$ is the lowest possible probability that the total primary demand is non-zero,
$U := \max \left\{ \left | \Omega_c \right |, \left | \Omega_s \right |,
    \left | \Omega_{a} \right | \right\}$
and $\varepsilon$ is the input parameter to Algorithm \ref{Alg_UCB}.
\end{theorem}


\begin{remark}
\textbf{Lower bound.} We note that the regret bound \eqref{eq:thm_regret} shown in Theorem \ref{main} is tight up to the logarithmic term.
More specifically, we can show that the regret is lower-bounded by
$\Omega \left( N M \hat{p} \sqrt{U T} \right)$.
The proof is based on constructing a problem instance
where this part of the regret is inevitable for any learning algorithm.

Consider the instance where the add-on space limit is $S=0$.
In addition, the primary demand $\alpha_{N+m}(\cdot)$ are zero,
and the add-on purchase probabilities $\beta_{N+m}(\cdot)$
are one, for all supportive products $m=\{1,\ldots,M\}$,
and all prices $p_{N+m} \in \Omega_s$.

In this instance, the optimal policy is to set the prices of all the supportive products at the highest price.
For simplicity, consider that there is only one price available for all supportive products, which is $\hat{p}$ as defined in Lemma \ref{lem_Lip}.

Now we can translate the problem into a collection of independent MAB problems.
In particular, for each core product, we obtain a regret lower bound $\Omega \left( M \hat{p} \sqrt{UT} \right)$.
Summing up the regret of all $N$ core products, we then obtain
the regret lower bound $\Omega \left( N M \hat{p} \sqrt{UT} \right)$. \Halmos

\end{remark}



The proof of Theorem \ref{main} is based on breaking down the total regret.

Before moving to the proof details,
we first introduce the notations that we need in the analysis.
Let $\overline{\alpha}_{n,t}(p_n)$, $\overline{\alpha}_{N+m,t}(p_{N+m})$,
$\overline{\beta}_{N+m,t}'(p_{N+m}')$, $\overline{\beta}_{N+m,t}(p_{N+m})$,
$c_{n,t}(p_n)$, $c_{N+m,t}(p_{N+m})$,
$c^{(a,1)}_{N+m,t}(p_{N+m}')$ and $c^{(a,2)}_{N+m,t}(p_{N+m})$
be the values of the corresponding parameters at the beginning of period $t$, respectively.
Then, with these notations, we define the collection of \emph{events} $\cE_t$,
where each parameter's empirical mean at the beginning of period $t$ is not in its confidence interval that is shown in Algorithm \ref{Alg_UCB}. Formally, we have
\begin{eqnarray*}
\cE_t &:=&
\left\{ \;
\bigcup_{n \in [N]} \bigcup_{p_n \in \Omega_c}
    \left| \overline{\alpha}_{n,t}(p_n) - \alpha_n(p_n) \right|
    > \frac{2\log t}{c_{n,t}(p_n)}
\;\right\}  \\
&& \bigcup \; \left\{ \;
\bigcup_{m\in[M]} \bigcup_{p_{N+m} \in \Omega_s}
    \left| \overline{\alpha}_{N+m,t}(p_{N+m}) -  \alpha_{N+m}(p_{N+m}) \right|
    > \frac{2\log t}{c_{N+m,t}(p_{N+m})}
\;\right\} \\
&&  \bigcup \;  \left\{ \;
\bigcup_{m\in[M]} \bigcup_{p_{N+m}' \in \Omega_a}
    \left| \overline{\beta}_{N+m,t}'(p_{N+m}') - \beta_{N+m,t}'(p_{N+m}') \right|
    > \frac{2 \log t}{c^{(a,1)}_{N+m,t}(p_{N+m}')}
\;\right\}  \\
&&  \bigcup \; \left\{ \;
\bigcup_{m\in[M]} \bigcup_{p_{N+m} \in \Omega_s}
    \left| \overline{\beta}_{N+m,t}(p_{N+m}) - \beta_{N+m}(p_{N+m}) \right|
    > \frac{2 \log t}{c^{(a,2)}_{N+m,t}(p_{N+m})}
\;\right\}.
\end{eqnarray*}

%
%

Conditioning on events $\cE_{t(\tau)}$ for all episodes,
we break down the total regret into two major parts.
Let $t(\tau)$ denote the starting period of episode $\tau$.
Let $n(\tau)$ be the total number of episodes from period $t=1$ to $T$, and
$\ell(\tau)$ the length of episode $\tau$, i.e., the total number of periods in episode $\tau$.
Specifically, we have
\begin{eqnarray}
\text{Regret}(T) &=&
\mathbf{E} \left [\sum_{t=1}^T  \mathcal{R}^* - \mathcal{R}(\Pi_t) \right ] \nonumber \\
&=& \mathbf{E} \left [ \sum_{\tau=1}^{n(\tau)} \left( \mathcal{R}^*- \mathcal{R}(\Pi_\tau) \right ) \cdot \ell(\tau) \right ] \nonumber  \\
&=&
\mathbf{E} \left [ \sum_{\tau=1}^{n(\tau)} \bE \left[ \left( \mathcal{R}^*- \mathcal{R}(\Pi_\tau) \right ) \cdot \ell(\tau) \mid \cE_{t(\tau)} \right] \cdot
\mathbf{P} \left[  \cE_{t(\tau)} \right] \right ] \nonumber \\
&+&  \mathbf{E} \left [ \sum_{\tau=1}^{n(\tau)} \bE \left[ \left( \mathcal{R}^*- \mathcal{R}(\Pi_\tau) \right ) \cdot \ell(\tau) \mid \cE_{t(\tau)}' \right ] \cdot
\mathbf{P} \left[ \cE_{t(\tau)}' \right] \right].
\label{eqn_regb}
\end{eqnarray}



In the following, we bound the first term in \eqref{eqn_regb} using Lemma \ref{reg_lem_2}
and Lemma \ref{reg_lem_3}, and bound the second term using Lemma \ref{reg_lem_4}.
Theorem \ref{main} then follows by summing up the two parts.

\begin{lemma} \label{reg_lem_2}
The expected length of the episode $\tau$ that starts with period $t$ is
upper-bounded by $t$, namely,
\begin{equation*}
    \mathbf{E} [\ell(\tau)] \leq t(\tau).
\end{equation*}
\end{lemma}

\textit{Proof.}
By definition, the associated counters for primary demand, i.e.,
${c_{n}(\cdot)}$ and ${c_{N+m}(\cdot)}$ always increase by $1$ in each period,
and the associated counters for add-on purchases, i.e.,
$c^{(a')}_{N+m}(\cdot)$ and $c^{(a)}_{N+m}(\cdot)$
could increase $0,1,\ldots, N$ in each period, which depends on the total number of core products purchased in the period.
In addition, to obtain an upper bound on the length of an episode, it suffices to consider only one counter that is associated with the primary demand.
W.L.O.G., consider the counter for product $1$ with its price determined at the beginning of episode $\tau$, namely, $t(\tau)$.
Then we know that the value of this counter is at most $t(\tau)-1$, and
the value will be doubled after another $t(\tau)-1$ periods.
By the description of Algorithm \ref{Alg_UCB}, episode $\tau$ starts in period $t(\tau)$ and terminates in no more than $t(\tau)-1$ periods.
Therefore, we have $\mathbf{E} [\ell(\tau)] \leq t(\tau)$. \Halmos

\begin{lemma} \label{reg_lem_3}
    Given the algorithm shown in Algorithm \ref{Alg_UCB}, we have
    \begin{equation} \label{eq_lem3}
        \mathbf{E} \left [ \sum_{\tau=1}^{n(\tau)} \bE \left[ \left( \mathcal{R}^*- \mathcal{R}(\Pi_\tau) \right ) \cdot \ell(\tau) \mid \cE_{t(\tau)} \right] \cdot
        \mathbf{P} \left[  \cE_{t(\tau)} \right] \right ] \leq K_1,
    \end{equation}
    where $K_1$ a constant that is independent of $T$.
\end{lemma}


\textit{Proof.}
By definition, $\cE_{t}$ is the union of a collection of events.

Specifically, for each core product $n\in\{1,\ldots,N\}$ and price $p_n \in \Omega_c$,
given the value of $t$ and $c_{n,t}(p_n)$, by the Chernoff-Hoeffding inequality, we have
$$\mathbf{P} \left\{
\left| \overline{\alpha}_{n,t}(p_n) - \alpha_n(p_n) \right| > \frac{2\log t}{c_{n,t}(p_n)}
\right\}  \leq \frac{2}{t^4}. $$
Take the union for all possible values of $c_{n,t}(p_n)$ from $1$ to $t$.
By the union bound,  we obtain
$$\mathbf{P} \left\{
\left| \overline{\alpha}_{n,t}(p_n) - \alpha_n(p_n) \right| > \frac{2\log t}{c_{n,t}(p_n)}
\right\}  \leq \frac{2}{t^3}. $$

Similarly, given the value of $t$, for each supportive product $m\in\{1,\ldots,M\}$
and price $p_{N+m} \in \Omega_s$,
take the union for all possible values of $c_{N+m,t}(p_{N+m})$ from $1$ to $t$,
and we obtain
$$\mathbf{P} \left\{
\left| \overline{\alpha}_{N+m,t}(p_{N+m}) -  \alpha_{N+m}(p_{N+m}) \right|
 > \frac{2\log t}{c_{N+m,t}(p_{N+m})}
\right\}  \leq \frac{2}{t^3}. $$

For add-on purchases, the counters $c^{(a,1)}_{N+m,t}(\cdot)$
and $c^{(a,2)}_{N+m,t}(\cdot)$
range from $1$ to $Nt$.
Thus, for each supportive product $m\in\{1,\ldots,M\}$, we obtain
$$\mathbf{P} \left\{
 \left| \overline{\beta}_{N+m,t}'(p_{N+m}') - \beta_{N+m,t}'(p_{N+m}') \right|
 > \frac{2 \log t}{c^{(a,1)}_{N+m,t}(p_{N+m}')}
\right\}  \leq \frac{2 N}{t^3}, $$
for each add-on discount price $p_{N+m}' \in \Omega_a$,
and
$$\mathbf{P} \left\{
 \left| \overline{\beta}_{N+m,t}(p_{N+m}) - \beta_{N+m}(p_{N+m}) \right|
 > \frac{2 \log t}{c^{(a,2)}_{N+m,t}(p_{N+m})}
\right\}  \leq \frac{2 N}{t^3}, $$
for each add-on original price $p_{N+m} \in \Omega_s$.

Take a union of all these events in $\cE_t$, we have
$$\mathbf{P} \left [ \cE_t \right] \leq \frac{2(N+M+2MN)U}{t^3}.$$

In addition, conditional on event $\cE_{t(\tau)}$, we know that the regret in episode $\tau$ is
upper-bounded by $\mathcal{R}^*\cdot \ell(\tau)$.
Therefore, for each term on the LHS of \eqref{eq_lem3}, we have
\begin{align*}
    &\bE \left[ \left( \mathcal{R}^*- \mathcal{R}(\Pi_\tau) \right ) \cdot \ell(\tau) \mid \cE_{t(\tau)} \right] \cdot \mathbf{P} \left[  \cE_{t(\tau)} \right] \\
    \leq\; & \mathcal{R}^* \cdot \ell(\tau) \cdot \frac{2(N+M+2MN)U}{t(\tau)^3} \\
    \leq\; & \mathcal{R}^* \cdot \frac{2(N+M+2MN)U}{ t(\tau)^2} \\
    \leq\; & \mathcal{R}^* \cdot \frac{2(N+M+2MN)U}{ \tau^2},
\end{align*}
where the second inequality follows from Lemma \ref{reg_lem_2}, and the third inequality follows by $t(\tau) \geq \tau$.

Take the sum over $\tau$ from $1$ to $n(\tau)$.
Since $\sum_{\tau=1}^{n(\tau)} 1/{\tau^2} \leq \pi^2/6$,
we obtain the upper bound
$$\mathcal{R}^* \cdot \frac{(N+M+2MN)U\pi^2}{3},$$
which concludes the proof of Lemma \ref{reg_lem_3}. \Halmos

\begin{lemma} \label{reg_lem_4}
Given the algorithm shown in Algorithm \ref{Alg_UCB}, we have
\begin{equation} \label{eq_lem4}
    \mathbf{E} \left [ \sum_{\tau=1}^{n(\tau)} \bE \left[ \left( \mathcal{R}^*- \mathcal{R}(\Pi_\tau) \right ) \cdot \ell(\tau) \mid \cE_{t(\tau)}' \right ] \cdot
    \mathbf{P} \left[ \cE_{t(\tau)}' \right] \right]
    \leq K_2 \cdot \left[ N M\hat{p} \left( (1/\lambda) \cdot \sqrt{UT\log T}
        + \varepsilon\sqrt{T}  \right) \right],
\end{equation}
where $K_2$ is a constant that is independent of $T$.
\end{lemma}

\textit{Proof.} For each term on the LHS of \eqref{eq_lem4}, we relax probability
$\mathbf{P} \left [  \cE_{t(\tau)}' \right]$ to $1$ as an upper bound.
We then need to show that the regret is bounded, conditional on event $ \cE_{t(\tau)}' $
where the empirical mean of each associated parameter is within its confidence interval.

Let $\Pi^*$ be the optimal policy, and
$U_{\tau}(\Pi^*)$ the value of the objective function of the optimization problem \eqref{opt_problem} under policy $\Pi^*$ with UCB input parameters
$\widetilde{\alpha}_n(p_n)$, $\widetilde{\alpha}_{N+m}(p_{N+m})$,
$\widetilde{\beta}_{N+m}'(p'_{N+m})$ and $\widetilde{\beta}_{N+m}(p_{N+m})$,
as defined in Algorithm \ref{Alg_UCB}.
Let $U_{\tau}(\Pi_\tau)$ be the value of the objective function of the optimization problem \eqref{opt_problem} under policy $\Pi_{\tau}$ with the same UCB input parameters.

Since the value of the objective function of \eqref{opt_problem} is increasing in all the parameters. We know that conditional on $\cE_{t(\tau)}$, $U_{\tau}(\Pi^*)$ is an upper bound of $\mathcal{R}^*$, namely, the expected revenue of the optimal policy,
and $U_{\tau}(\Pi_\tau)$ is an upper bound of $\mathcal{R}(\Pi_\tau)$.
Therefore, we have
\begin{align} \label{ucb_ineq}
&\mathbf{E} \left [ \sum_{\tau=1}^{n(\tau)} \bE \left[ \left( \mathcal{R}^*- \mathcal{R}(\Pi_\tau) \right ) \cdot \ell(\tau) \mid \cE_{t(\tau)}' \right] \right] \nonumber \\
=\; & \mathbf{E} \left [ \sum_{\tau=1}^{n(\tau)} \bE \left[ \left(
 \mathcal{R}^*
 - U_\tau(\Pi^*) + U_\tau(\Pi^*)
 - U_\tau(\Pi_\tau) + U_\tau(\Pi_\tau)
 - \mathcal{R}(\Pi_\tau) \right ) \cdot \ell(\tau) \mid \cE_{t(\tau)}' \right] \right]\nonumber \\
 \leq\; & 4\hat{p}MN \varepsilon \sqrt{T}
  + \bE \left [ \sum_{\tau=1}^{n(\tau)} \bE \left[ \left( U_\tau(\Pi_\tau) - \mathcal{R}(\Pi_\tau) \right ) \cdot \ell(\tau) \mid \cE_{t(\tau)}' \right] \right].
\end{align}

The inequality follows because $ \mathcal{R}^* - U_\tau(\Pi^*) \leq 0$
and  $U_\tau(\Pi^*)- U_\tau(\Pi_\tau)$ is upper bounded by the approximation error of the FPTAS optimization subroutine.
Specifically, with parameter
$K=\lceil \sqrt{t(\tau)}{\varepsilon} \rceil$, we have
\begin{equation*}
U_{\tau}(\Pi^*) - U_{\tau}(\Pi_\tau)
\leq U_{\tau}(\Pi_\tau^*) -  U_{\tau}(\Pi_\tau)
= \frac{\hat{p}MN}{\lceil {\sqrt{t(\tau)}}/{\varepsilon} \rceil}
\leq {\hat{p}MN\varepsilon}/{\sqrt{t(\tau)}}.
\end{equation*}
By Lemma \ref{reg_lem_2}, we know $\ell(\tau) \leq t(\tau)$.
Therefore, we have
\begin{equation*}
\sum_{\tau=1}^{n(\tau)} \hat{p}MN\varepsilon/{\sqrt{t(\tau)}}
\leq \sum_{t=1}^T \hat{p}MN\varepsilon \cdot \left(2/\sqrt{t}\right)
\leq 4\hat{p}MN \varepsilon \sqrt{T}.
\end{equation*}

The second term in \eqref{ucb_ineq}, namely,
$\bE \left [ \sum_{\tau=1}^{n(\tau)} \bE \left[ \left( U_\tau(\Pi_\tau) - \mathcal{R}(\Pi_\tau) \right ) \cdot \ell(\tau) \mid \cE_{t(\tau)}' \right] \right]$,
describes the \emph{confidence} bound for the revenue of policy $\Pi_{\tau}$,

Let $p_{n,\tau}$, $p_{N+m,\tau}$, $p_{N+m,\tau}'$ and $I_{N+m,\tau}$
be the decisions of policy $\Pi_{\tau}$.

Conditional on event $\cE_{t(\tau)}'$, we obtain
\begin{eqnarray} \label{ineq_sens}
U_\tau(\Pi_\tau) - \mathcal{R}(\Pi_\tau)
\leq 2 (M+1) \hat{p} \cdot \sum_{n=1}^N \sqrt{\frac{2\log t(\tau)}{c_{n,t(\tau)}(p_{n,\tau})} }
+ 2 \hat{p} \cdot \sum_{m=1}^M \sqrt{\frac{2\log t(\tau)}{c_{N+m,t(\tau)}(p_{N+m,\tau})}}
 \nonumber\\
+ 2 N \hat{p} \cdot \sum_{m=1}^M \left[ I_{N+m,\tau} \cdot
    \sqrt{\frac{2\log t(\tau)}{c^{(a,1)}_{N+m,t(\tau)}(p_{N+m,\tau}')}}
+ \left(1- I_{N+m,\tau}\right) \cdot
    \sqrt{\frac{2\log t(\tau)}{c^{(a,2)}_{N+m,t(\tau)}(p_{N+m,\tau})}} \right].
\end{eqnarray}

%

To show the inequality, we observe that each of the four parts in \eqref{ineq_sens} in fact
corresponds to the revenue gap due to the over estimation of the associated demand parameters.
In addition, for each parameter, we know that the gap between its true mean and its UCB term is
$2\sqrt{\frac{2\log t(\tau)}{\text{counter}}}$ conditional on event $\cE_{t(\tau)}'$.

Specifically, we have that
\begin{itemize}
\item parameter $\alpha_n(p_{n,\tau})$ contributes to the revenue gap by at most
$(M+1)\hat{p}\cdot 2 \sqrt{\frac{2\log t(\tau)}{c_{n,t(\tau)}(p_{n,\tau})}}$;
\item parameter $\alpha_{N+m}(p_{N+m,\tau})$ contributes to the revenue gap by at most
$\hat{p}\cdot 2 \sqrt{\frac{2\log t(\tau)}{c_{N+m,t(\tau)}(p_{N+m,\tau})}}$;
\item parameter $\beta_{N+m}'(p_{N+m,\tau}')$ contributes to the revenue gap by at most
$N\hat{p}\cdot 2 \sqrt{\frac{2\log t(\tau)}{c^{(a,1)}_{N+m,t(\tau)}(p_{N+m,\tau}')}}$
if $I_{N+m}=1$, and nothing if $I_{N+m}=0$;
\item parameter $\beta_{N+m}(p_{N+m,\tau})$ contributes to the revenue gap by at most
$N\hat{p}\cdot 2 \sqrt{\frac{2\log t(\tau)}{c^{(a,2)}_{N+m,t(\tau)}(p_{N+m,\tau})}}$
if $I_{N+m}=0$, and nothing if $I_{N+m}=1$.
\end{itemize}

Given inequality \eqref{ineq_sens}, we take a sum over $\tau$ on both sides and obtain the following.

For parameter $\alpha_n(\cdot)$, we have
\begin{eqnarray} \label{ineq_bound2}
&&\bE \left[ \sum_{\tau=1}^{n(\tau)} \ell(\tau) \cdot \sqrt{\frac{2\log t(\tau)}{c_{n,t(\tau)}(p_{n,\tau})}} \mid \cE_{t(\tau)}' \right]
\leq \bE \left[ \sum_{t=1}^{T} \sqrt{\frac{4 \log t(\tau)}{c_{n,t}(p_{n,t})}}
\mid \cE_{t(\tau)}' \right] \nonumber\\
&\leq& \bE \left[ \sum_{i=1}^{\zeta(c)} \sum_{j=1}^{c_{n,T}(q_c^{i})} \sqrt{\frac{4\log T}{j}} \mid \cE_{t(\tau)'} \right] =
2 \sqrt{\log T} \; \bE \left[ \sum_{i=1}^{\zeta(c)} \sum_{j=1}^{c_{n,T}(q_c^{i})} \sqrt{\frac{1}{j}} \mid \cE_{t(\tau)}' \right] \nonumber\\
&\leq&  2 \sqrt{\log T} \;  \left[
\sum_{i=1}^{\zeta(c)} 2 \sqrt{c_{n,T}(q_c^{i})} \mid \cE_{t(\tau)}' \right]
\leq 2 \sqrt{\log T} \; \bE \left[
2\sqrt{UT} \mid \cE_{t(\tau)}' \right]
=  4 \sqrt{ UT \log T}.
\end{eqnarray}

In the first inequality, with abuse of notation, we use $p_{n,t}$ to denote the price decision of product $n$ in period $t$. The inequality follows due to the fact that the counter's value in period $\tau$ is no larger than $2 \cdot t(\tau)$.
The second inequality follows from the relaxation of $\log t(\tau)$ to $\log T$, and an alternative way of counting $c_{n,t}(p_{n,t})$ from $t=1$ to $T$.
The third inequality follows because of the fact that $\sum_{j=1}^{J}\sqrt{1/j}\leq 2\sqrt{J}$.
The last inequality follows from the Cauchy-Schwarz inequality
since we know $\zeta(c):= |\Omega_c| \leq U$ and $\sum_{i=1}^{\zeta(c)} c_{n,T}(q_c^{i}) \leq T$.

Following the same analysis, we can show similar bounds for parameters $\alpha_{N+m}(\cdot)$, $\beta_{N+m}'(\cdot)$ and $\beta_{N+m}(\cdot)$.
Notice that in developing the bounds for $\beta_{N+m}'(\cdot)$ and $\beta_{N+m}(\cdot)$,
we need to have the values of the associated counters $c^{(a,1)}_{N+m,T}(\cdot)$ and $c^{(a,2)}_{N+m,T}(\cdot)$ to be non-zero for at least one of the prices.
Hence, we need to multiple the bound by $1/\lambda$, where $\lambda$ denotes the lowest probability that the total primary demand is non-zero.

Now given \eqref{ineq_sens} and \eqref{ineq_bound2},
we have
\begin{equation*}
\mathbf{E} \left[ \sum_{\tau=1}^{n(\tau)} \ell(\tau) \cdot \left(  U_\tau(\Pi_\tau)- \mathcal{R}(\Pi_\tau) \right] \mid \cE_{t(\tau)} \right] \leq
\left[ 2(M+1) N  + 2 M + \frac{4N M}{\lambda} \right]  \hat{p} \cdot 4\sqrt{UT\log T}.
\end{equation*}

Put the bound back to \eqref{ucb_ineq}, and we obtain
\begin{equation*}
\bE \left[ \bE \left[ \sum_{\tau=1}^{n(\tau)} \ell(\tau) \cdot
\left( \mathcal{R}^*- \mathcal{R}(\Pi_\tau) \right] \mid \cE_{t(\tau)}' \right]
\cdot \mathbf{P} \left [ \cE_{t(\tau)}' \right] \right]
\leq 4\hat{p}MN\varepsilon\sqrt{T}
+ \frac{8 N M}{\lambda}  \hat{p} \cdot 4 \sqrt{UT\log T}.
\end{equation*}

This concludes the proof of Lemma \ref{reg_lem_4}. \Halmos

The proof of Theorem \ref{main} then follows by combining the results in Lemma \ref{reg_lem_2}, \ref{reg_lem_3}, and \ref{reg_lem_4}.

%% file: Discussion.tex
\section{Discussion} \label{sec_dis}

Given that the revenue management problem with add-on discounts is a new model,
we discuss in this section several variants of the optimization problem
for different practical scenarios.
In particular, we discuss how the changes of the underlying model assumptions would affect the formulation of the problem,
 as well as the optimization and learning algorithms.

\subsection{Assumption on independent demand}

In optimization problem \eqref{opt_problem},
we have assumed that all the demand parameters (both primary demand and add-on purchase) are independent across different products,
which means that all the demand functions, $\alpha_n(p_n)$, $\alpha_{N+m}(p_{N+m})$, $\beta_{N+m}'(p_{N+m}')$ and $\beta_{N+m}(p_{N+m})$, are dependent only on the price of the corresponding product itself.
Alternatively, we can model demand using discrete choice models, e.g. the MNL model and other similar variants.

Consider the demand assumption that each customer can purchase at most one core product.
Let $\mathbf{p}_c$ denote the vector of prices of the core products, i.e., $\mathbf{p}_c:= (p_1, \ldots, p_N)$, and $\alpha_n(\mathbf{p}_c)$ denote the purchase probability for core product $n$ given price vector $\mathbf{p}_c$.
Given a choice model, we have $\sum_{n=1}^N \alpha_n(\mathbf{p}_c) \leq 1$.
However, for supportive products, as one customer can purchase multiple supportive products at the same time, it is not appropriate to model demand functions
$\alpha_{N+m}(\cdot)$, $\beta_{N+m}'(\cdot)$ and $\beta_{N+m}(\cdot)$
using choice models (because with choice models, e.g. MNL, each customer can select at most one product).
The revenue management problem with a choice model for the core products can now be formulated as:


\begin{equation} \label{opt_problem2}
\begin{aligned}
\max _{p_n, p_{N+m}, p_{N+m}', I_{N+m}} \quad & \sum_{n=1}^{N}
\alpha_n(\mathbf{p}_c) \cdot p_n
+\sum_{m=1}^{M}\alpha_{N+m}(p_{N+m}) \cdot p_{N+m}  \\
& \quad \quad + \left [ \sum_{n=1}^{N}\alpha_n(\mathbf{p}_c) \right ] \cdot \sum_{m=1}^{M} \left [I_{N+m} \cdot \beta_{N+m}^{'}(p_{N+m}^{'}) \cdot p_{N+m}^{'} \right ]  \\
& \quad \quad + \left [ \sum_{n=1}^{N}\alpha_n(\mathbf{p}_c) \right ] \cdot \sum_{m=1}^{M} \left [ \left ( 1- I_{N+m} \right )\cdot \beta_{N+m}(p_{N+m}) \cdot p_{N+m} \right ]\\
\textrm{s.t.} \quad & \sum_{m=1}^{M} I_{N+m} \le S ,  \\
& p_{N+m}' < p_{N+m} \text{ for } m=1,\ldots, M,  \\
& p_n \in \Omega_c \text{ for } n=1,\ldots, N,\\
& p_{N+m} \in \Omega_s,  \;
p_{N+m}' \in \Omega_{a},  \; I_{N+m}\in \{0,1\} \text{ for } m=1,\ldots, M.
\end{aligned}
\end{equation}

Given the new formulation \eqref{opt_problem2}, we cannot apply Algorithm \ref{Alg_sub} to solve the corresponding offline optimization problem.
Although we can use Part 1 of Algorithm \ref{Alg_sub} to approximate the revenue function $\mathcal{R}_s(\cdot)$ for the supportive products, we cannot apply Part 2 of Algorithm \ref{Alg_sub} to solve the master problem
$\max \sum_{n=1}^{N}\alpha_n(\mathbf{p}_c) p_n
+ \mathcal{R}_s(\sum_{n=1}^{N} \alpha_n(\mathbf{p}_c))$
using the same dynamic programming approach.
In the following, we discuss the solutions to the updated optimization problem in two cases:
1) when the number of core products $N$ is small 2and ) when $N$ is large.
We also discuss how to solve the joint learning and optimization problem in both cases.

When $N$ is small relative to the computational resource, we can solve the optimization problem by enumerating the value of $\sum_{n=1}^{N}\alpha_n(\mathbf{p}_c) p_n+ \mathcal{R}_s(\sum_{n=1}^{N}\alpha_n(\mathbf{p}_c))$
over all possible price vectors $\mathbf{p}_c$.
Correspondingly, in the joint learning and optimization problem, we can consider the choice model as a black-box, and use this enumerating solution as a subroutine.
Specifically, we can model each possible price vector $\mathbf{p}_c$ as an arm, and learn
the values of $\sum_{n=1}^{N}\alpha_n(\mathbf{p}_c)p_n$ and $\sum_{n=1}^{N}\alpha_n(\mathbf{p}_c)$ separately using a UCB-based algorithm similar to Algorithm \ref{Alg_UCB}.
Moreover, we can show that such a learning algorithm can converge to the optimal policy, with a slower convergence rate,
as the algorithm's regret is now proportional to the number of all possible price vectors, i.e., $O \left( |\Omega_c|^N \right)$.


When $N$ is large and we can no longer consider the choice model as a black-box,
we have to resort to heuristics based on neighborhood searching to obtain near-optimal solutions. In addition, we need to explicitly incorporate the choice model into the learning algorithm.
In this case, if we assume the underlying choice model to be MNL, we can use the
existing learning algorithms for MNL models (e.g., \cite{rusmevichientong2010dynamic}, \cite{agrawal2016near},
\cite{agrawal2017thompson}, \cite{agrawal2019mnl}, or \cite{MX2019}) to handle the learning task in our problem.


\subsection{Assumption on add-on demand function}

We assume in the optimization problem \eqref{opt_problem} that the probability of an add-on purchase under discount price $\beta_{N+m}'(\cdot)$ depends only on discount price $p_{N+m}'$, rather than on both $p_{N+m}'$ and $p_{N+m}$.
This assumption is justified by the following two observations from practice.
First, many supportive products, like video games, have \emph{suggested retail prices} from the industry. For example, in the US, the prices of video games are regularly set at $\$59.99$.
This fixed price, rather than the offered price $p_{N+m}$, can be considered as a reference point for customers. Therefore, it suffices to consider only the discount price $p_{N+m}'$ in estimating demand $\beta_{N+m}'(\cdot)$.
Second, in practice, add-on discounts are usually shown as a \emph{limited time offer} as a way of triggering a customer's intention to purchase.
Therefore, in the purchase dynamics, it is common that customers only consider whether or not to take discount price $p_{N+m}'$, instead of going back and comparing the discount price with the original price $p_{N+m}$.

Alternatively, one can assume that $\beta_{N+m}'(\cdot)$ depends both on the discount price $p_{N+m}'$ and the original price $p_{N+m}$.
In this case, we can update the formulation of the offline optimization problem as follows.

\begin{equation} \label{opt_problem3}
\begin{aligned}
\max _{p_n, p_{N+m}, p_{N+m}', I_{N+m}} \quad & \sum_{n=1}^{N} \alpha_n(p_n)p_n
+ \sum_{m=1}^{M}\alpha_{N+m}(p_{N+m})p_{N+m} \\
& \quad \quad + \left [ \sum_{n=1}^{N}\alpha_n(p_n) \right ] \cdot \sum_{m=1}^{M} \left [I_{N+m} \cdot \beta_{N+m}'(p_{N+m}',p_{N+m}) \cdot p_{N+m}^{'} \right ]  \\
& \quad \quad + \left [ \sum_{n=1}^{N}\alpha_n(p_n) \right ] \cdot \sum_{m=1}^{M} \left [ \left ( 1- I_{N+m} \right )\cdot \beta_{N+m}(p_{N+m}) \cdot p_{N+m} \right ]\\
\textrm{s.t.} \quad & \sum_{m=1}^{M} I_{N+m} \le S ,  \\
& p_{N+m}' < p_{N+m} \text{ for } m=1,\ldots, M . \\
& p_n \in \Omega_c \text{ for } n=1,\ldots, N,\\
& p_{N+m} \in \Omega_s,  \;
p_{N+m}' \in \Omega_{a},  \; I_{N+m}\in \{0,1\} \text{ for } m=1,\ldots, M.
\end{aligned}
\end{equation}

In this updated formulation, we replace the $\beta_{N+m}'(p_{N+m}')$ in \eqref{opt_problem} with $\beta_{N+m}'(p_{N+m}',p_{N+m})$. Note that this modification will not change the framework of the optimization subroutine shown in Algorithm \ref{Alg_sub}, and hence we can still apply Algorithm \ref{Alg_sub} to solve the offline problem.
More specifically, in Part 1.(b) of Algorithm \ref{Alg_sub}, we simply update the procedure to $$\max_{p_{N+m}\in \Omega_s, p_{N+m}'\in \Omega_{a}, p_{N+m}' < p_{N+m} } \alpha_{N+m}(p_{N+m})p_{N+m}+ \gamma \beta_{N+m}'(p_{N+m}')p_{N+m}',$$
and the algorithm's complexity stays unchanged.

For the joint learning and optimization problem, we can adopt Algorithm \ref{Alg_UCB}
with a simple modification of the counters associated with function $\beta_{N+m}'(\cdot)$.
Following a similar regret analysis procedure to that in Section \ref{sec_learn},
we obtain regret
$$\mathcal{O} \left( N M \hat{p} \left( 1/\lambda \sqrt{U^2 T \log T}
+ \varepsilon \sqrt{T}  \right) \right), $$
where the original term $U$ shown in Theorem \ref{main} is now replaced by $U^2$.

\subsection{Assumption on Bernoulli demand}

We assume in our model that all the demand parameters $\alpha_n(p_n)$, $\alpha_{N+m}(p_{N+m})$, $\beta_{N+m}'(p_{N+m}')$ and $\beta_{N+m}(p_{N+m})$
are represented by Bernoulli random variables.
In fact, the model can be extended to other types of demand parameters
as long as we can obtain similar concentration results for the learning algorithm, as shown in Lemma \ref{reg_lem_3} and Lemma \ref{reg_lem_4}.
In our analysis, we use the Chernoff-Hoeffding inequality to obtain the concentration results for Bernoulli random variables.
We refer interested readers to \cite{bubeck2013bandits} for further discussions on the concentration results for other types of random variables, such as normal, Poisson, exponential and all bounded distributions, which all belong to the family of sub-exponential distributions.


In this paper, our major focus is the add-on discount structure in the revenue management problem, and hence we skip the detailed discussion on the assumptions of the underlying demand random variables, as well as the corresponding regret analysis.
In fact, relaxing the Bernoulli assumptions will only affect the construction of the UCB terms, and the framework of our learning and optimization algorithm will still apply. In addition, in practice, one may simply remove the $\min(1,\cdot)$ term in the UCB to handle other types of demand parameters.

%% file: Numerical.tex
\section{Numerical Experiments} \label{sec_num}

We present in this section the results of our numerical experiments.
We conduct the experiments with the real-world data we collect from Tmall.com, which is an online e-commerce platform operated in China by Alibaba Group.
The data provide the transaction history from a popular video gaming brand's official online store at Tmall.com.
In the experiments, we first use the data to estimate the demand-price relationships of different products as the \emph{ground truth}.
Then we test the performance of algorithm $\ucb$ in different settings with varying levels of add-on discount effects and add-on space limits.
The experiment results not only validate the performance guarantee of the learning algorithm $\ucb$, as shown in Theorem \ref{main}, but also illustrate the advantages of using the add-on discount strategy in practice.

\subsection{Experiment settings}

The data provide the detailed transaction records from the video gaming brand's online store at Tmall.com during the period from October 2017 to July 2019.
The store mainly sells video game consoles, video games and accessories.
We observe from the data that the major sales are from three video game consoles and twenty video games. Therefore, we set $N=3$ and $M=20$ in all the experiments.

We use the data to calculate the \emph{hourly} arrival rate, i.e, the number of customers per hour, as the demand for each of the selected video game consoles (core products) and video games (supportive products).
In addition, for each product, given its demand under different prices, we use linear models to estimate the demand functions, i.e., $\alpha_n(p_n)$, $\alpha_{N+m}(p_{N+m})$ and $\beta_{N+m}(p_{N+m})$, as the ground truth.
More specifically, for each $m\in \{ 1,\ldots,M \}$, we estimate $\alpha_{N+m}(p_{N+m})$ and $\beta_{N+m}(p_{N+m})$ separately:
if the game is purchased together with a game console, we then count the transaction as add-on demand $\beta_{N+m}(p_{N+m})$;
and if the game is purchased without any game console, we count the transaction as primary demand $\alpha_{N+m}(p_{N+m})$.
The details of the estimated coefficients of functions $\alpha_n(p_n)$, $\alpha_{N+m}(p_{N+m})$ and $\beta_{N+m}(p_{N+m})$ are provided in the Appendix.
We note that the linear demand assumption is only used for estimating the ground truth, and is not known to the learning algorithm.

Since the online store does not implement any add-on discount strategy, we cannot estimate the add-on demand function $\beta_{N+m}'(p_{N+m}')$ from the transaction data directly.
Instead, we generate these functions based on $\beta_{N+m}(p_{N+m})$ by making different assumptions about the level of the add-on discount effect.
Given the intuition that for each video game, its add-on demand should be higher than its primary demand under the same price, we consider the following three cases in our experiments:
\begin{itemize}
\item \textbf{Low} add-on discount effect, where
$\beta_{N+m}'(\cdot) = 2 \cdot \beta_{N+m}(\cdot)$ for all $m=1,\ldots,M$;
\item \textbf{Medium} add-on discount effect, where
$\beta_{N+m}'(\cdot) = 3 \cdot \beta_{N+m}(\cdot)$ for all $m=1,\ldots,M$;
\item \textbf{High} add-on discount effect, where
$\beta_{N+m}'(\cdot) = 4 \cdot \beta_{N+m}(\cdot)$ for all $m=1,\ldots,M$.
\end{itemize}

Given the total number of games $M=20$, we consider three possible values for the space limits, i.e., the total number of add-on discounts the retailer can offer at most,
which is $S \in \{4,6,8\}$. In total, we test $9$ cases ($3$ levels of add-on discount effect $\times$ $3$ space limits) in our experiments.

We also modify the prices of different products that are used in practice.
First, since the sale prices of video game consoles are much higher than those of the video games, we subtract the unit cost, which we assume to be $3,000$ CNY, from the sale price of each game console, in order to obtain prices of the same level in the objective function \eqref{opt_problem}.
With this modification, we can consider the objective value as the total profit rather than the total revenue.
Second, for simplicity, we round the prices that end with $9$ or $99$ to the nearest ten or hundred.
We set the price sets of different products as follows.
For video game consoles, we have $p_n \in \Omega_c = \{200,400,600,800\}$.
For video games, we have $p_{N+m} \in \Omega_s = \{80,100,120,140,160\}$ and
$p_{N+m}' \in \Omega_a = \{80,100,120,140\}$.
All the prices are in CNY.
Note that price value $160$ is removed from $\Omega_{a}$,
as it never makes a feasible add-on discount.

When running algorithm $\ucb$, we set the approximation error to be $\varepsilon=0.1$ for the optimization subroutine. This approximation error is also used for
calculating the revenue of the optimal policy with Algorithm \ref{Alg_sub}.
In addition, in constructing the confidence intervals, we add an additional multiplier, which we fix to be $2^{-3}$, to all the UCB terms to enhance the algorithm's efficiency.
The reasons for adding this multiplier are further discussed in \cite{RVR2014}.
Moreover, we note that each period in our experiment corresponds to one hour in the real world. This is consistent with our calculation of demand, which is defined as the hourly arrival rate. We also note that $365 \times 24 = 8760$ periods in our experiments correspond to the time of a year in the real-world.

\subsection{Result Analysis}

We aim to answer the following questions in analyzing the results of our numerical experiments.

\begin{itemize}
\item How does algorithm $\ucb$ perform in different scenarios, in terms of the algorithm's rate of convergence to the optimal policy that knows the true demand functions?
\item What is the optimality gap, namely, the difference in total revenue, between the optimal policy that uses add-on discounts (i.e., $S>0$), and the optimal policy that does not use add-on discounts (i.e., $S=0$), when both optimal policies know the true demand functions from the beginning?
\item How long does it take for algorithm $\ucb$ to achieve a better performance, in terms of total revenue (profit), than the optimal policy that does not use add-on discounts?
\end{itemize}

We summarize the experiment results under different test scenarios in Table \ref{tab_num}.

In the first part of the table, we demonstrate the performance of algorithm $\ucb$.
For each test scenario, we run the algorithm a total of $100$ times,
and then calculate the average regret, as defined in \eqref{eq:def_regret}, up to period $T=168$ (one week), period $T=672$ (one month), period $T=2016$ (three months) and period $T=8760$ (one year), respectively.
We display the average regret in percentage, which is given by
\begin{align}
\frac{\text{Regret}(T)}{\mathcal{R}^* \cdot T} = 1 -
\frac{\sum_{t=1}^T  \mathbf{E} \left[ \mathcal{R}(\Pi_t) \right] }{\mathcal{R}^* \cdot T}.
\end{align}

In the second part of Table \ref{tab_num}, we answer the second question by displaying
the difference of total revenue (in percentage) between the optimal policy that uses add-on discounts and the optimal policy that does not use add-on discounts.
For simplicity, we call the first policy the \emph{optimal (add-on) policy} and the second policy the \emph{optimal no-add-on policy}.
Let $\mathcal{R}_{0}^*$ be the revenue of the optimal no-add-on policy.
The optimality gap percentage is given by
$\left(\mathcal{R}^*/\mathcal{R}_{0}^* - 1\right)$.

In the last column of Table \ref{tab_num}, we show the number of periods it takes for algorithm $\ucb$ to surpass the revenue of the optimal no-add-on policy.

\begin{figure}[htbp]
\begin{center}
\includegraphics[width=14cm]{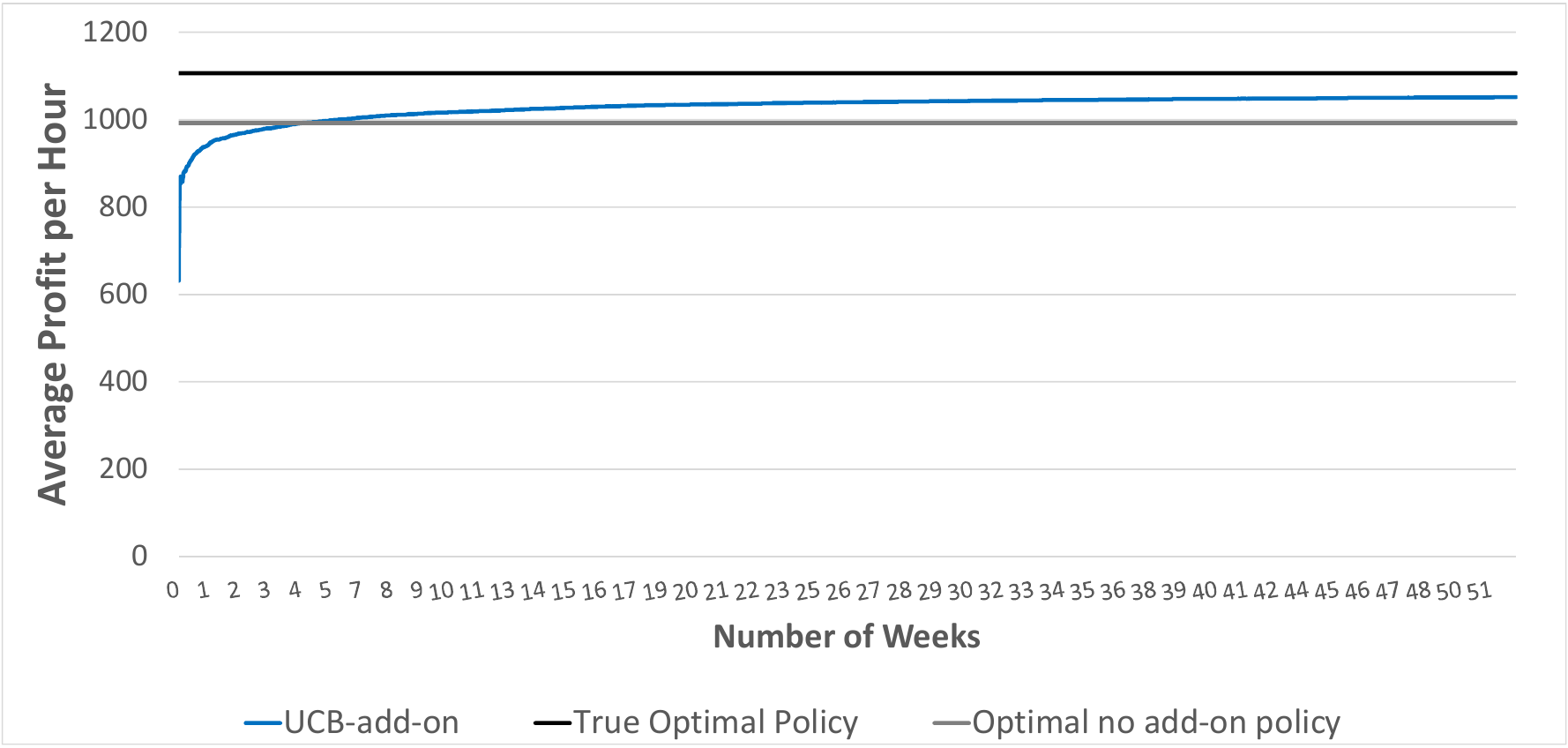}
\caption{Performance of algorithm $\ucb$ in the test case for $S=6$ with medium add-on discount effect.}
\label{f_con}
\end{center}
\end{figure}

To visualize the performance of algorithm $\ucb$ in comparison to the two optimal policies,
we plot out the accumulative revenue of our algorithm
as a function of the real-world time in Figure \ref{f_con}.
Specifically, the results are from the test case where $S=6$ and the add-on effect is medium.
The plot also shows the comparisons between our algorithm and the other two optimal policies.

\begin{table}[htbp]
  \centering
  \caption{Summary of experiment results under different test scenarios.}
    \begin{tabular}{|r|c|c|c|c|c|c|}
    \hline
    \multicolumn{7}{|c|}{\textbf{Low} add-on discount effect :
    $\beta_{N+m}'(\cdot) = 2\cdot \beta_{N+m}(\cdot) $} \bigstrut\\
    \hline
    \multicolumn{1}{|c|}{} & \multicolumn{4}{c|}{Expected average regret percentage} & Optimality gap of  & Time to beat optimal  \bigstrut\\
\cline{1-5}    \multicolumn{1}{|c|}{Time} & 1 week & 1 month & 3 months & 1 year & optimal no add-on policy & no add-on policy \bigstrut\\
    \hline
    $S=4$ & 14.10\% & 10.60\% & 8.00\% & 5.50\% & 4.30\% & \textit{1.2 years} \bigstrut\\
    \hline
    $S=6$ & 11.00\% & 9.20\% & 7.30\% & 3.90\% & 5.60\% & \textit{0.5 year} \bigstrut\\
    \hline
    $S=8$ & 16.60\% & 11.40\% & 7.80\% & 5.10\% & 6.30\% & \textit{0.5 year} \bigstrut\\
    \hline
    \multicolumn{7}{|c|}{\textbf{Medium} add-on discount effect:
    $\beta_{N+m}'(\cdot) = 3\cdot \beta_{N+m}(\cdot) $ } \bigstrut\\
    \hline
    \multicolumn{1}{|c|}{} & \multicolumn{4}{c|}{Expected average regret percentage} & Optimality gap of  & Time to beat optimal  \bigstrut\\
\cline{1-5}    \multicolumn{1}{|c|}{Time} & 1 week & 1 month & 3 months & 1 year & optimal no add-on policy & no add-on policy \bigstrut\\
    \hline
    $S=4$ & 14.80\% & 11.30\% & 7.20\% & 5.30\% & 8.70\% & \textit{2 months} \bigstrut\\
    \hline
    $S=6$ & 17.10\% & 11.90\% & 7.70\% & 4.80\% & 11.50\% & \textit{1 month} \bigstrut\\
    \hline
    $S=8$ & 12.50\% & 10.10\% & 6.80\% & 4.50\% & 12.80\% & \textit{1/4  month} \bigstrut\\
    \hline
    \multicolumn{7}{|c|}{\textbf{High} add-on discount effect:
    $\beta_{N+m}'(\cdot) = 4\cdot \beta_{N+m}(\cdot) $} \bigstrut\\
    \hline
    \multicolumn{1}{|c|}{} & \multicolumn{4}{c|}{Expected average regret percentage} & Optimality gap of  & Time to beat optimal  \bigstrut\\
\cline{1-5}    \multicolumn{1}{|c|}{Time} & 1 week & 1 month & 3 months & 1 year & optimal no add-on policy & no add-on policy \bigstrut\\
    \hline
    $S=4$ & 11.50\% & 8.30\% & 6.50\% & 4.50\% & 13.20\% & \textit{6 days} \bigstrut\\
    \hline
    $S=6$ & 16.40\% & 11.20\% & 7.00\% & 4.40\% & 17.20\% & \textit{6 days} \bigstrut\\
    \hline
    $S=8$ & 14.10\% & 9.30\% & 6.60\% & 4.10\% & 19.30\% & \textit{2 days} \bigstrut\\
    \hline
    \end{tabular}%
    \label{tab_num}%
\end{table}%

We make the following observations from the results shown in Table \ref{tab_num} and Figure \ref{f_con}.

First, we observe that algorithm $\ucb$ can efficiently converge to the optimal policy in all test scenarios.
The regret (in percentage) shrinks to $10\%$ within one-month time in all the tests. In addition, the figure validates the algorithm's convergence rate $\mathcal{O}(\sqrt{T})$, as shown in Theorem \ref{main}.

Second, for the optimality gap between the two optimal policies, we observe that the gap increases when $S$ becomes larger and when the add-on discount effect becomes stronger.
The results are reasonable because in both cases, the revenue of the optimal add-on policy increases, while the revenue of the optimal no add-on policy stays the same.
Moreover, we observe a non-negligible optimality gap: even in the modest setting where $S=4$ and the add-on discount effect is low, the gap is $4.3\%$.
Such comparison results demonstrate the advantages of using the add-on discount strategy.

Third, from the comparisons between algorithm $\ucb$ and the optimal no-add-on policy, we observe that the time for algorithm $\ucb$ to beat the optimal-add-on policy decreases as the space limit or the discount effect increases.
This is consistent with our observations from the optimality gap comparisons.
More importantly, the results reassure the benefits of using the add-on discount strategy even when the retailer has no prior knowledge of all the demand parameters.
As we show in Figure \ref{f_con}, where the $x$-axis depicts the real time in weeks, and the $y$-axis depicts the average hourly revenue (i.e., revenue per period),
the learning algorithm can quickly outperform the optimal no-add-on policy in around four weeks.

%% file: Conclusion.tex
\section{Conclusions and Future Research Directions} \label{sec_con}

In this paper, we study a revenue management problem with add-on discounts, which is motivated by the unique structure between core products (video game consoles) and supportive products (video games).
We note that although the add-on discount strategy has been used in the industry,
it has not been formally studied in the Operations Management literature, and our work fills this gap between theory and practice.
In particular, we develop an optimization formulation of the revenue management problem, and provide an FPTAS algorithm that can approximately solve the optimization problem to any desired accuracy.
Moreover, we study the problem in the online setting where the demand functions of different products are unknown. We propose a UCB-based algorithm to solve the online problem, and show that the algorithm can obtain a tight regret bound.

This paper also provides useful managerial insights and strategical guidance for retailers.
In principle, the add-on discount strategy offers more flexibility for product promotions,
and retailers can increase their revenue (and profit) by adopting this strategy so as to incentivize customers to purchase more items.
However, in practice, the lack of past experience and the uncertainty of customer's demand could hold retailers back from implementing the strategy.
In our numerical experiments, which are based on the real-world data we collect from Tmall.com, we show that the retailer can expect a revenue (profit) increase of $5\%$ to $20\%$ by using add-on discounts.
More importantly, in the more practical setting where the retailer has no prior knowledge of the demand information, we show that the retailer can obtain a long-term increase in revenue (profit) by using the add-on discount strategy while learning the demand parameters on the fly.
These numerical results demonstrate the efficacy of using data-driven approaches in revenue management.



We conclude the paper by pointing out several interesting future research directions.

First, our model motivates a more general add-on setting where discounts are offered \emph{two-way}.
More specifically, in this paper, we categorize the products into core products and supportive products, and assume that the retailer can only offer add-on discounts on supportive products.
In the more general setting, given two selected sets of products, we assume that the retailer can offer add-on discounts to any set of products.
The challenges of studying this general add-on discount model include:
1) developing a good formulation of the problem;
2) analyzing the offline optimization problem;
and 3) designing the online learning algorithm.

Second, building on the results of this paper, it worth exploring another innovative revenue management strategy called \emph{share-for-discounts}.
In share-for-discounts, customers can collect bonus points by sharing the information of certain products with their friends.
Once the bonus points reach some threshold, a customer can get discounts on the shared products as rewards.
By using this strategy, retailers can reach more potential customers through a customer's personal social network.
Therefore, how to design a good data-driven policy for the share-for-discounts strategy would be another interesting research direction.


%% file: Appendix.tex
\newpage
\setcounter{page}{1}
\setcounter{section}{0}

\section*{Appendix - Parameters in Numerical Testing}

In Table \ref{tab_ap1} and \ref{tab_ap2}, we show our estimations of the coefficients (intercepts and slopes) for the demand functions of three video game consoles and twenty video games, respectively, using the real-world transaction data.

Note that the demand for each product under all allowable prices
is always between $[0,1]$,
and thus can be interpreted as the mean of a Bernoulli random variable.

\begin{table}[htbp]
  \centering
  \caption{Parameters for $\alpha_n(\cdot)$}
    \begin{tabular}{|r|r|r|}
    \hline
    $n$     & Intercept & Slope \bigstrut\\
    \hline
    1     & 0.975 & -7.25E-04 \bigstrut\\
    \hline
    2     & 0.27  & -2.00E-04 \bigstrut\\
    \hline
    3     & 1.15  & -8.50E-04 \bigstrut\\
    \hline
    \end{tabular}%
  \label{tab_ap1}%
\end{table}%

\begin{table}[htbp]
  \centering
  \caption{Parameters for $\alpha_{N+m}(\cdot)$ and $\beta_{N+m}(\cdot)$}
    \begin{tabular}{|r|r|r|r|r|}
    \hline
    \multicolumn{1}{|c|}{\multirow{2}[4]{*}{$m$}} & \multicolumn{2}{c|}{$\alpha_{N+m}(\cdot)$} & \multicolumn{2}{c|}{$\beta_{N+m}(\cdot)$} \bigstrut\\
\cline{2-5}    \multicolumn{1}{|c|}{} & Intercept & Slope & Intercept & Slope \bigstrut\\
    \hline
    1     & 0.085 & -4.38E-04 & 0.050 & -2.50E-04 \bigstrut\\
    \hline
    2     & 0.353 & -1.81E-03 & 0.208 & -1.04E-03 \bigstrut\\
    \hline
    3     & 0.097 & -5.00E-04 & 0.057 & -2.88E-04 \bigstrut\\
    \hline
    4     & 0.073 & -3.75E-04 & 0.043 & -2.13E-04 \bigstrut\\
    \hline
    5     & 0.044 & -2.25E-04 & 0.027 & -1.38E-04 \bigstrut\\
    \hline
    6     & 0.260 & -1.34E-03 & 0.153 & -7.63E-04 \bigstrut\\
    \hline
    7     & 0.029 & -1.50E-04 & 0.017 & -8.75E-05 \bigstrut\\
    \hline
    8     & 0.024 & -1.25E-04 & 0.015 & -7.50E-05 \bigstrut\\
    \hline
    9     & 0.066 & -3.38E-04 & 0.038 & -1.88E-04 \bigstrut\\
    \hline
    10    & 0.013 & -6.25E-05 & 0.008 & -3.75E-05 \bigstrut\\
    \hline
    11    & 0.243 & -1.25E-03 & 0.143 & -7.13E-04 \bigstrut\\
    \hline
    12    & 0.015 & -7.50E-05 & 0.008 & -3.75E-05 \bigstrut\\
    \hline
    13    & 0.063 & -3.25E-04 & 0.037 & -1.88E-04 \bigstrut\\
    \hline
    14    & 0.129 & -6.63E-04 & 0.077 & -3.88E-04 \bigstrut\\
    \hline
    15    & 0.095 & -4.88E-04 & 0.057 & -2.88E-04 \bigstrut\\
    \hline
    16    & 0.019 & -1.00E-04 & 0.012 & -6.25E-05 \bigstrut\\
    \hline
    17    & 0.019 & -1.00E-04 & 0.012 & -6.25E-05 \bigstrut\\
    \hline
    18    & 0.316 & -1.63E-03 & 0.187 & -9.38E-04 \bigstrut\\
    \hline
    19    & 0.241 & -1.24E-03 & 0.142 & -7.13E-04 \bigstrut\\
    \hline
    20     & 0.019 & -1.00E-04 & 0.012 & -6.25E-05 \bigstrut\\
    \hline
    \end{tabular}%
    \label{tab_ap2}%
\end{table}%